\documentclass{article} 
\usepackage{iclr2025_conference,times}

\usepackage{amsmath,amsfonts,bm}

\def\eqref#1{equation~\ref{#1}}

\def\1{\bm{1}}

\DeclareMathAlphabet{\mathsfit}{\encodingdefault}{\sfdefault}{m}{sl}
\SetMathAlphabet{\mathsfit}{bold}{\encodingdefault}{\sfdefault}{bx}{n}

\usepackage{hyperref}
\usepackage{url}
\usepackage{wrapfig,lipsum,booktabs}
\usepackage{adjustbox}
\usepackage{xspace}
\usepackage{enumitem}  
\usepackage{multirow}
\usepackage{colortbl}
\usepackage{tikz}

\usepackage{tocloft}
\usepackage{titlesec}
\usepackage{etoolbox}
\usepackage{titletoc}

\newcommand\DoToC{%
  \startcontents
  \printcontents{}{1}{\noindent \textbf{\Large{Table of Contents in Appendix}}\vskip3pt\vskip5pt}
  \vskip3pt\vskip5pt
}

\definecolor{cblue}{RGB}{33, 95, 154}
\definecolor{cgreen}{RGB}{59, 125, 35}

\title{Teach Multimodal LLMs to Comprehend Electrocardiographic Images}

\author{$^\clubsuit$Ruoqi Liu\thanks{Equal Contribution}, $^\clubsuit$Yuelin Bai\footnotemark[1], $^\lozenge$Xiang Yue, $^\clubsuit$Ping Zhang\\
$^\clubsuit$The Ohio State University,
$^\lozenge$Carnegie Mellon University\\
}

\newcommand{\model}{\texttt{PULSE}\xspace}
\newcommand{\data}{\texttt{ECGInstruct}\xspace}
\newcommand{\eval}{\texttt{ECGBench}\xspace}

\iclrfinalcopy 
\begin{document}

\maketitle

\vspace{-1.0cm}
\begin{center}
    \url{https://aimedlab.github.io/PULSE/}
\end{center}
\vspace{0cm}

\begin{abstract}

The electrocardiogram (ECG) is an essential non-invasive diagnostic tool for assessing cardiac conditions. Existing automatic interpretation methods suffer from limited generalizability, focusing on a narrow range of cardiac conditions, and typically depend on raw physiological signals, which may not be readily available in resource-limited settings where only printed or digital ECG images are accessible. Recent advancements in multimodal large language models (MLLMs) present promising opportunities for addressing these challenges. However, the application of MLLMs to ECG image interpretation remains challenging due to the lack of instruction tuning datasets and well-established ECG image benchmarks for quantitative evaluation. To address these challenges, we introduce \data, a comprehensive ECG image instruction tuning dataset of over one million samples, covering a wide range of ECG-related tasks from diverse data sources. Using \data, we develop \model, an MLLM tailored for ECG image comprehension. In addition, we curate \eval, a new evaluation benchmark covering four key ECG image interpretation tasks across nine different datasets. Our experiments show that \model sets a new state-of-the-art, outperforming general MLLMs with an average accuracy improvement of 15\% to 30\%. This work highlights the potential of \model to enhance ECG interpretation in clinical practice.

\end{abstract}

\begin{figure*}[!hb]
    \centering
    \vspace{-5pt}
    \includegraphics[width=0.96\textwidth]{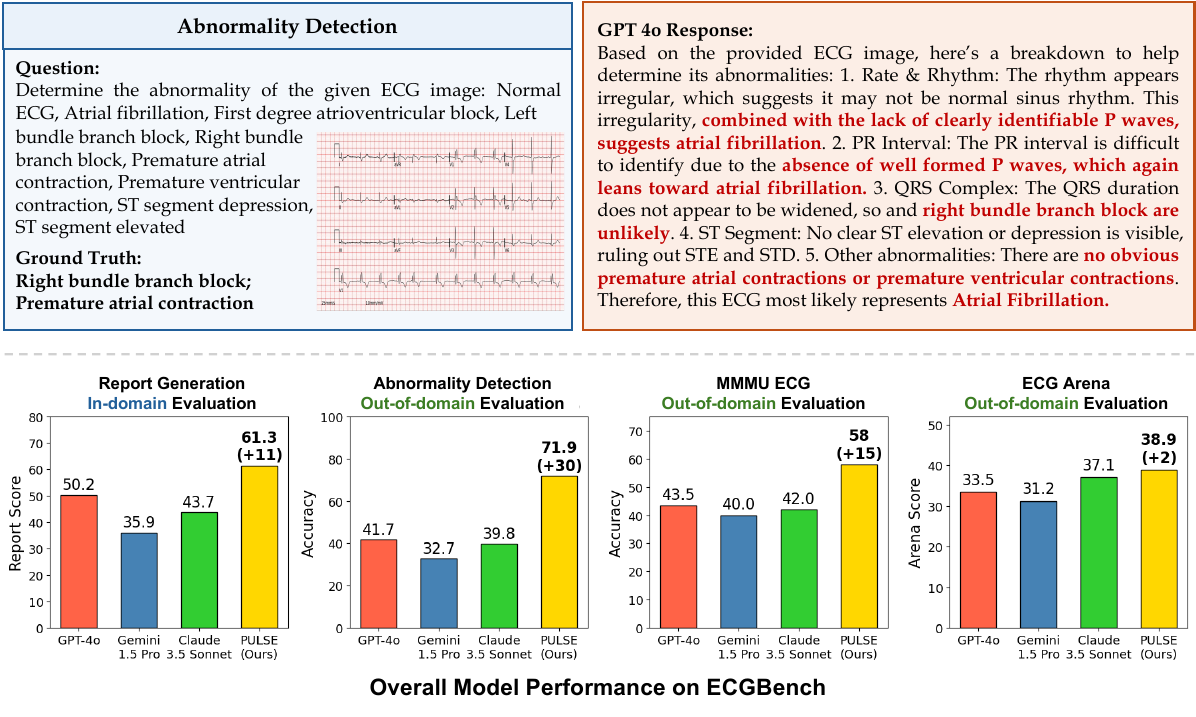}
    \vspace{-5pt}
    \caption{The proposed \model demonstrates superior performance across multiple \textcolor{cblue}{in-domain} and \textcolor{cgreen}{out-of-domain} datasets on our constructed \eval compared with advanced proprietary MLLMs (e.g., GPT-4o).  Notably, the proprietary MLLMs often fail to accurately interpret ECG images, generating well-structured and contextually relevant responses but ultimately incorrect (with errors highlighted in red) compared to the ground truth diagnosis.}
    \label{fig:overview}
    \vspace{-5pt}
\end{figure*}

\section{Introduction}
The electrocardiogram (ECG) is an essential tool in diagnosing cardiovascular diseases due to its non-invasive, cost-effective, and widely accessible nature for assessing cardiac function. While some approaches have been proposed for automatic ECG diagnosis~\citep{hannun2019cardiologist, ribeiro2020automatic, hughes2021performance}, these are primarily designed for classification tasks with limited cardiac conditions, often lacking generalizability. Moreover, they typically treat ECG data as \textit{time-series physiological signals}, which may not always be available, particularly in resource-constrained settings where only \textit{printed or digital images} are accessible~\citep{sangha2022automated, sangha2023detection}.

Recent advancements in multimodal large language models (MLLMs) have shown impressive success across vision-language tasks, offering new possibilities for addressing the limitations of traditional ECG models. However, applying MLLMs to ECG interpretation is not straightforward. As illustrated in Fig. \ref{fig:overview}, current MLLMs, such as GPT-4o~\citep{openai2024gpt4o}, often provide responses that appear correct and contextually relevant but are ultimately inaccurate in interpreting ECG images. This highlights the need for specialized MLLMs for ECG image interpretation.

Developing MLLMs for ECG images faces several challenges. First, no large-scale ECG image datasets are currently available as most ECG datasets contain only raw signal data, which needs to be synthesized into digital images. Second, there is a lack of instruction tuning datasets for ECG images. Large high-quality instruction tuning datasets, which are crucial for MLLM development, need to be curated from scratch for ECG-related tasks. Finally, evaluation is just as critical as model development, yet no established benchmark exists for assessing MLLM performance in ECG image interpretation. A well-defined benchmark is essential for both quantifying model performance and identifying areas for future improvement.

In this paper, we tackle these challenges by introducing \data, the first large-scale ECG image instruction tuning dataset containing over one million ECG image-text samples. \data is characterized by: 1) realistic image synthesis that replicates artifacts commonly seen in paper-based ECGs, 2) a diverse range of ECG-related tasks refined with insights from clinical experts, and 3) data sourced from distinct geographic regions. Leveraging \data, we develop \model, an MLLM for ECG image comprehension. To evaluate \model, we present \eval, a comprehensive evaluation benchmark covering four ECG image interpretation tasks across nine different datasets. \eval includes repurposed tasks (e.g., abnormality detection) from existing datasets, and newly curated, more challenging tasks using real-world ECG images.

Evaluated on \eval, \model sets a new state-of-the-art, significantly outperforming proprietary MLLMs across all benchmarks with an average accuracy gain of 15\% to 30\% compared to GPT-4o on out-of-domain datasets (Fig. \ref{fig:overview}). Ablation experiments demonstrate the importance of incorporating diverse data sources and ECG instruction tasks into the training data. A case study and discussion further illustrate the model's effectiveness in ECG image interpretation.

To summarize, our main contributions are as follows,
\begin{itemize}[leftmargin=*]
    \item \textbf{Problem.} We investigate the capabilities of MLLMs in ECG image interpretation and evaluate their performance across various downstream tasks. To the best of our knowledge, this is the first study focused on assessing MLLMs in image-based ECG interpretation.
    \item \textbf{Dataset.} We construct \data, a large-scale ECG image instruction tuning dataset consisting of a wide range of ECG-related tasks, serving as a valuable resource for fine-tuning MLLMs for ECG image interpretation.
    \item \textbf{Model.} We develop \model, a new MLLM tailored for ECG image interpretation. The model achieves state-of-the-art performance, outperforming both proprietary and open-source MLLMs.
    \item \textbf{Evaluation.} We establish \eval, a comprehensive benchmark for evaluating ECG image interpretation, which includes diverse evaluation tasks, both real-world and synthesized images. 
\end{itemize}


\begin{figure*}[!t]
    \centering
    \includegraphics[width=0.99\textwidth]{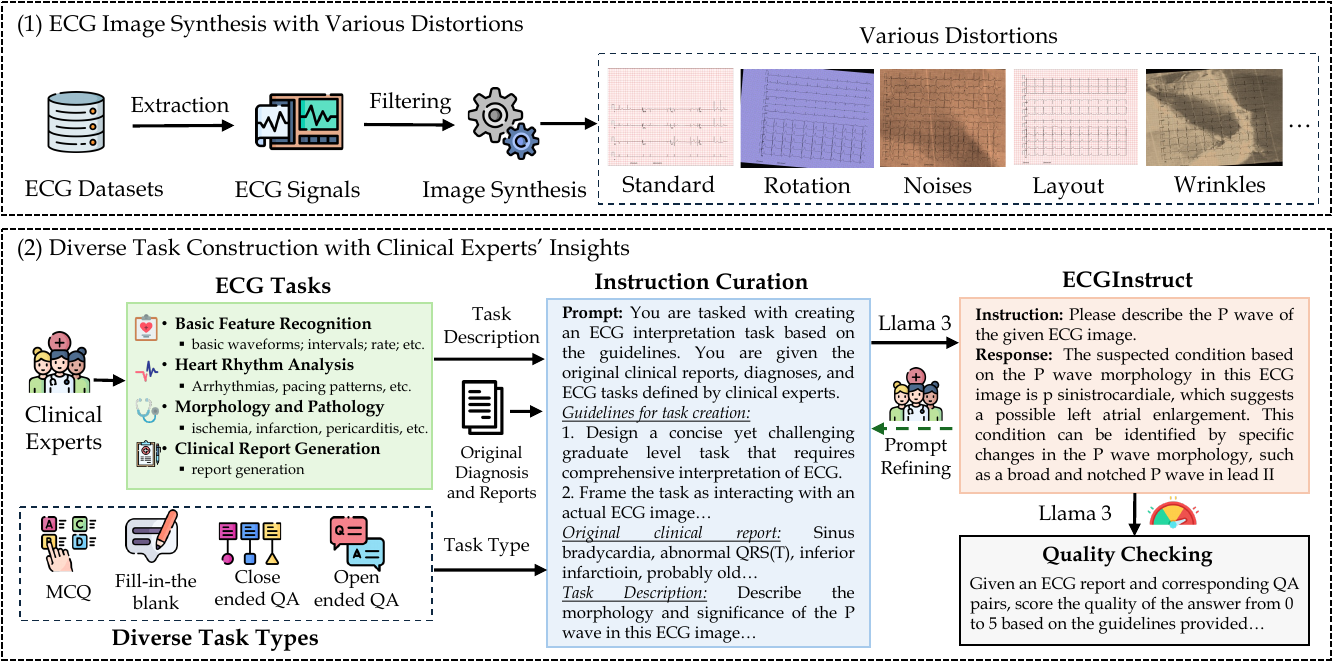}
    \caption{\data: a list of diverse and large-scale instruction tuning datasets for ECG image interpretation. (1) ECG images are synthesized from raw signal recordings with various distortions that mimic real-world printed ECG images. (2) \data is curated based on clinician-defined ECG-related tasks, original diagnosis and clinical reports, and diverse task types. Additional quality checking is applied to filter lower-scored instructions.}
    \label{fig:ecg-instruct-overview}
\end{figure*}

\section{ECGInstruct: Teach MLLMs to Comprehend ECG Images}
We aim to curate a list of multifaceted instruction tuning datasets for ECG analysis that are featured by 1) realistic image synthesis resembling the artifacts in paper ECGs, 2) diverse types of ECG-related tasks with clinical experts' insights, and 3) different data sources from distinct geographical regions. We show the construction of \data in Fig. \ref{fig:ecg-instruct-overview} and data summary in Table \ref{tab:ecg-instruct}.

\begin{table*}[!t]
\begin{center}
\small
\begin{tabular}{llll}
\toprule
Source Dataset & Task & Type & \# Samples \\
\midrule
PTB-XL~\citep{wagner2020ptb} & Feature          & Close/Open/Fill/MCQ & 30K  \\
       & Rhythm        & Close/Open/Fill/MCQ & 36K  \\
       & Morphology         & Close/Open/Fill/MCQ & 67K  \\
       & Report & Open & 16K \\
\midrule
ECG-QA~\citep{oh2024ecg} & Feature          & Close & 40K  \\
       & Rhythm        & Close & 9K   \\
       & Morphology         & Close & 90K  \\
\midrule
MIMIC-IV-ECG~\citep{gow2023mimicivecg}  & Feature          & Close/Open/Fill/MCQ & 29K  \\
       & Rhythm        & Close/Open/Fill/MCQ & 115K \\
       & Morphology         & Close/Open/Fill/MCQ & 169K \\
       & Report & Open & 487K \\
\midrule
CODE-15\%~\citep{ribeiro2021code} & Feature          & Close & 22K  \\
       & Rhythm        & Close & 14K  \\
       & Morphology         & Close & 31K  \\ \midrule
Total (\data) &           &  & 1.2M  \\
\bottomrule
\end{tabular}
\end{center}
\caption{Summary of \data. Feature: basic feature recognition, Rhythm: heart rhythm analysis, Morphology: morphology and pathology identification, Report: clinical report generation. Close: close-ended QA, Open: open-ended QA, Fill: fill-in-the-blank, MCQ: multi-choice QA. The full table of data statistics is provided in Appendix Table \ref{tab:ecg-instruct-stat-all}.}
\label{tab:ecg-instruct}
\end{table*}

\subsection{ECG Image Synthesis with Various Distortions} To enhance the robustness and real-world applicability of our model, we synthesize ECG images mimicking common artifacts found in paper ECGs. We adopt an ECG image synthesis tool~\citep{shivashankara2024ecg} that provides various imperfections such as grid line interference, creases, wrinkles, paper rotations, etc. By including these synthesized artifacts, we aim to train models that can effectively interpret ECGs in less-than-ideal conditions, as often encountered in clinical settings. More details are provided in Appendix \ref{sec:image_synthesis}. 

\subsection{ECG-related Tasks with Clinical Experts' Insights} To construct a comprehensive set of ECG-related tasks, we consulted domain experts to curate diverse and clinically relevant tasks covering four different categories. Each category is designed to address specific aspects of ECG interpretation and analysis, including (1) basic feature recognition (see examples in Appendix Fig. \ref{fig:train_data_basic}), (2) heart rhythm analysis (see examples in Appendix Fig. \ref{fig:train_data_rhythm}), (3) morphology and pathology identification (see examples in Appendix Fig. \ref{fig:train_data_morph_path}) and (4) clinical report generation (see examples in Appendix Fig. \ref{fig:train_data_report}). Basic feature recognition (e.g., interval or segment, etc.) forms the foundation of ECG interpretation, enabling the model to grasp essential cardiac parameters. Heart rhythm analysis (e.g., arrhythmias, conduction abnormalities, etc.) and morphology and pathology identification (e.g., wave shape, pathological conditions, etc.) are more advanced and critical aspects of ECG analysis, ensuring that the model can detect and classify complex conditions accurately. Lastly, clinical report generation mirrors the process of healthcare professionals synthesizing a comprehensive interpretation of an ECG. By incorporating clinical experts' insights, we encourage the model to learn the practical skills required in a clinical context.

\subsection{Diverse Types of Tasks and Data Sources}
Based on the original diagnoses and clinical reports from the existing ECG datasets, we curate diverse types of tasks including multi-choice questions, fill-in-the-blank, close-ended QA, and open-ended QA. 
This variety of task types not only enhances the model's versatility but also mimics the diverse cognitive processes involved in real-world ECG interpretation. By incorporating these varied task types, we aim to develop a more robust and adaptable model capable of handling a wide spectrum of ECG-related queries and analyses.

To ensure broad applicability and generalizability, we collect ECG data from four different sources across geographically distinct regions: 1) PTB-XL~\citep{wagner2020ptb}: a Germany-based, publicly available repository; (2) MIMIC-IV-ECG~\citep{gow2023mimicivecg}: a large set of ECGs for patients who appear in the MIMIC-IV Clinical Database from Beth Israel Deaconess Medical Center in Boston~\citep{johnson2023mimic}; 
3) CODE-15\%~\citep{ribeiro2021code}: an ECG dataset from a central ECG repository from Minas Gerais, Brazil under the clinical outcomes in digital electrocardiology (CODE) study~\citep{ribeiro2019tele}; 4) ECG-QA~\citep{oh2024ecg}, a question answering dataset for ECGs that is constructed based on PTB-XL~\citep{wagner2020ptb}. This diverse geographical representation enhances the model's ability to generalize across different populations and healthcare systems, accounting for potential variations in ECG patterns and interpretations across regions.

\subsection{Data Synthesizing at Scale} Since large-scale annotation of ECG features is extremely expensive and time-consuming, we develop an automatic data synthesizing pipeline to address this data scarcity issue. We utilized clinical reports from PTB-XL and MIMIC-IV-ECG as initial seed data and leveraged an advanced LLM (i.e., Llama-3-70B-Instruct) for data synthesis.  Building upon the expert-in-the-loop process and diverse data resources described in the previous sections, we synthesized a substantial volume of ECG-related instructions and corresponding responses. These were based on expert-provided examples and real-world scenarios, with the specific prompts used in this process detailed in the Appendix \ref{sec:prompt}.
For datasets lacking comprehensive reports, such as CODE-15\%, we manually constructed diverse templates to transform the existing data into an instruction-response format.

\subsection{Quality Control}
To guarantee the quality of generated instructions and corresponding responses, we apply an independent LLM as a judge to evaluate and score the content. This process involves several steps: 1) initial generation: instructions and responses are first generated using our primary model; 2) evaluation criteria: we establish a set of evaluation criteria including the instruction relevance, clarity, answerability of the responses, etc; 3) LLM judge and scoring: an independent LLM (Llama 3~\citep{meta2024llama3}) is used as a judge to assess each instruction-response pair against established criteria and assign scores (see prompt in Appendix Fig. \ref{fig:prompt_instruction_score}); 4) feedback loop: low-scoring items are flagged for human expert review and potential revision or removal; 5) iterative refinement: based on the scoring patterns and human expert input, we continually refine our instruction generation process. By combining automated LLM evaluation with human expert oversight, we create a robust system for maintaining and improving the quality of our instruction-response pairs. 

\subsection{Training}
Our model architecture closely follows that of LLaVA~\citep{liu2024improved,liu2024llavanext}, adapting it for ECG image analysis. We use a vision encoder to process ECG images and a large language model as the text decoder, connected via a projection layer.
We organize the data into three components: the image, the instructions, and the outputs. The instruction is query or task related to the ECG image and the output is the expected response or prediction base on the image and instruction. We place the image at the beginning of each conversation, serving as the visual grounding for the entire dialogue.
During training, we freeze the parameters of the vision encoder while updating the parameters of the projection layer and the language model using an autoregressive training objective, where we mask all the tokens belonging to the image and the instruction.

\begin{figure*}[!t]
    \centering
    \includegraphics[width=0.99\textwidth]{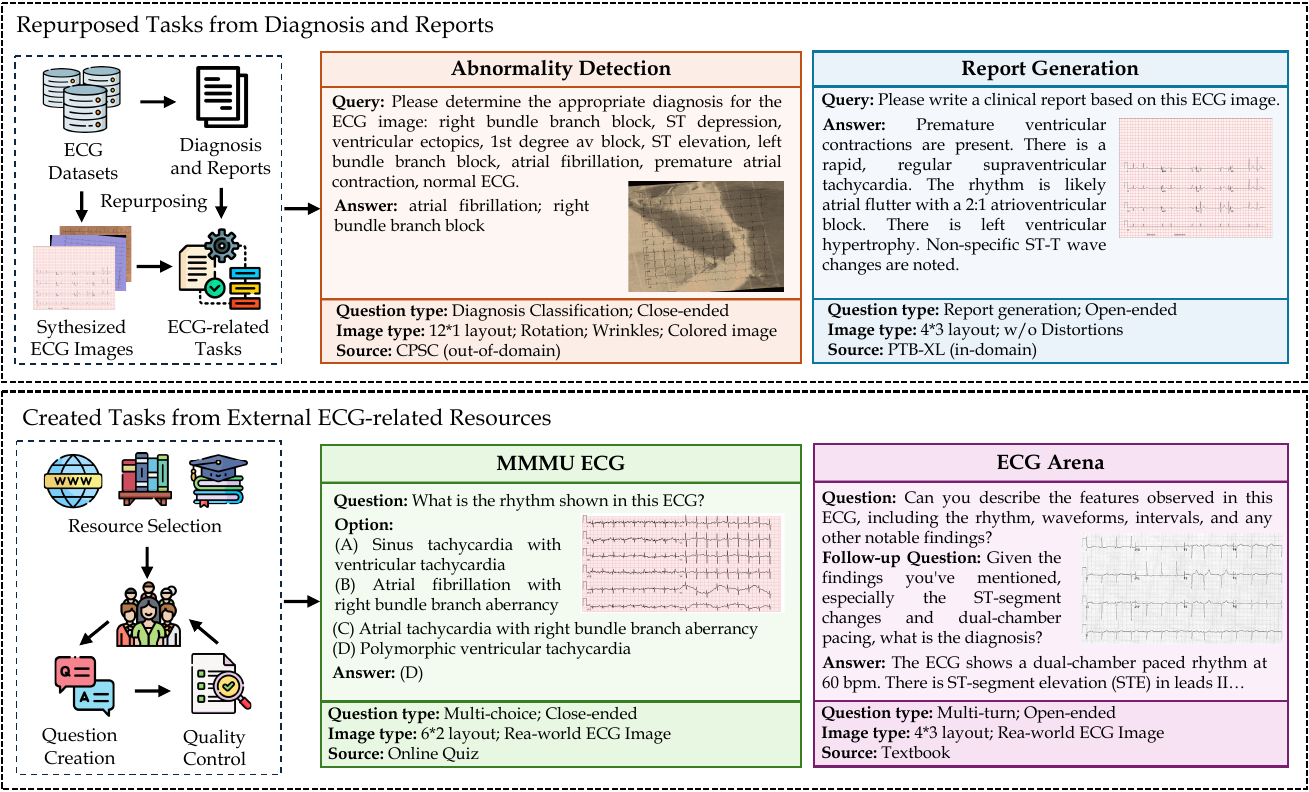}
    \caption{The data curation process for \eval. There are four key tasks involved: (1) two repurposed tasks (abnormality detection and report generation) derived from existing ECG datasets, where ECG images are synthesized from raw signals, and queries/answers are extracted based on diagnostic and clinical reports; (2) Two newly developed tasks using external resources, where ECG images and associated questions and answers are collected and generated from real-world sources.}
    \label{fig:ecg-bench}
\end{figure*}


\section{ECGBench}
In this section, we present ECG-Bench (Fig. \ref{fig:ecg-bench}), a comprehensive benchmark for evaluating MLLMs on ECG image interpretation. Our benchmark contains both repurposed tasks from six existing datasets and newly created tasks from external resources. Table \ref{tab:evaluation_datasets} shows the details of each evaluation dataset. We introduce the detailed evaluation task curation process below. 


\begin{table*}[!t]
\begin{center}
\adjustbox{max width=0.9\linewidth}{
\begin{tabular}{lllll}
\toprule
Evaluation Dataset & Task & Type & \# Samples & In-Domain? \\ \midrule
PTB-XL Super & Abnormality Detection & Close-ended & 2,082 & YES \\
PTB-XL Report & Report Generation & Open-ended & 500 & YES \\
CODE-15\% & Abnormality Detection & Close-ended & 1,400 & YES \\
ECG-QA & Abnormality Detection & Close-ended & 1,317 & YES \\ \midrule
CPSC 2018 & Abnormality Detection & Close-ended & 2,061 & NO \\
CSN & Abnormality Detection & MCQ (8-option) & 1,611 & NO \\
G12EC & Abnormality Detection & MCQ (8-option) & 2,026 & NO \\
MMMU ECG & Multimodal Understanding & MCQ (4-option) & 200 & NO \\
ECG Arena & Multi-turn Conversation & Open-ended & 50 & NO \\ \bottomrule
\end{tabular}}
\end{center}
\caption{Overview of evaluation datasets in \eval. This collection contains both in-domain and out-of-domain problems across four key tasks with diverse answer types.}
\label{tab:evaluation_datasets}
\end{table*}

\subsection{Evaluation Task Curation}
\textbf{Abnormality Detection.} This task focuses on detecting cardiac abnormalities using ECG images. We curate this task by repurposing six existing ECG datasets: three in-domain datasets: PTB-XL (Super)~\citep{wagner2020ptb}, CODE-15\%~\citep{ribeiro2021code}, ECG-QA~\citep{oh2024ecg}, and three out-of-domain datasets: CPSC 2018~\citep{liu2018open}, CSN~\citep{zheng2020optimal, zheng202012} and G12EC~\citep{liu2018open}. For all datasets, we first synthesize images using raw signals and then curate queries based on the original diagnosis and reports. For datasets with fewer than 10 diagnostic labels, we curate close-ended questions. For those with more labels, we construct multi-choice questions with 8 options, including the original diagnosis and randomly sampled negative labels.

\textbf{Report Generation.} This task involves generating detailed reports for given ECG images. We benchmark using 500 randomly selected reports from the test set of PTB-XL, which contains high-quality ECG reports written and validated by cardiologists. Similarly, the ECG images are synthesized from the raw signals. For the ground truth reports written in non-English (PTB-XL is a Germany-based dataset), we translate the reports into English before the evaluation. 

\textbf{MMMU ECG.} Inspired by MMMU~\citep{yue2024mmmu}, a widely adopted evaluation benchmark for MLLMs, we manually curated an ECG version with 200 multi-choice questions with the help of medical school students. The curation process involved three key steps: 
(1) \textbf{Resource Selection:} We gathered ECG materials from diverse and reliable sources such as ECG textbooks, clinical case reports from medical journals, and widely used online ECG learning materials. This ensures the comprehensiveness and quality of collected ECG examples and interpretations.
(2) \textbf{Question Creation and Collection:} Five medical school students with basic knowledge of ECG were recruited for this task. They extracted existing questions from the collected resources. For ECG images accompanied only by clinical interpretations, the annotators created questions based on these interpretations. Additionally, they formulated new questions drawing from their expertise, ensuring a balance between various ECG interpretation aspects (e.g., rhythm analysis, morphology assessment, clinical interpretation).
(3) \textbf{Quality Control:} To maintain high standards, we implemented a quality control process. In particular, Each question underwent review by at least two other annotators, checking for accuracy and clarity. An independent reviewer cross-checked the final images, questions, and corresponding answers against the original sources to ensure fidelity to the source material. Any discrepancies or ambiguities were resolved during this process. 

\textbf{ECG Arena.} To assess the model's instruction-following ability in ECG comprehension, we developed ECG Arena, inspired by MT-Bench ~\citep{zheng2024judging} and Arena-hard ~\citep{chiang2024chatbot} used in general LLM chat evaluations. We manually curated 50 multi-turn ECG-related questions, focusing on open-ended interactions. The data curation process for ECG Arena, like MMMU ECG, involves three main steps: resource selection, question creation, and quality control. The key distinction is that MMMU ECG focuses on multiple-choice questions, whereas ECG Arena involves more complex, flexible multi-turn, open-ended questions. Each follow-up question is contingent on the initial question and its response, making the process more challenging and reflective of real-world applications. Since multi-turn conversations are rare in existing sources, this posed significant challenges during data curation. To address this, annotators created such conversations by referencing original clinical interpretations and ECG images. The questions are designed to feel natural and simulate a real clinical setting (e.g., the first question may ask about basic findings from the image, followed by a question about potential clinical causes or diagnoses based on those findings).

\subsection{Evaluation Metrics}
\textbf{Abnormality Detection:} We use macro AUC, macro F1, and hamming loss (HL) for multi-label datasets, and accuracy for others. \textbf{Report Generation:} We employ GPT-4o as a judge, evaluating reports based on rhythms, waveform, and diagnosis, with a maximum score of 100 points (see evaluation prompt in Appendix Fig. \ref{fig:prompt_report_eval}).
\textbf{MMMU ECG:} We use accuracy as the primary metric, with systematic, rule-based evaluation pipelines to ensure consistent scoring.
\textbf{ECG Arena:} GPT-4o assesses model performance by comparing generated responses with ground truth answers, considering accuracy, completeness, and instruction adherence, with a maximum score of 100 points (see evaluation prompt in Appendix Fig. \ref{fig:prompt_arena_eval}). More evaluation details are provided in the Appendix \ref{sup:eval_details}.
\section{Experiments}

\subsection{Methods for Comparison}
In order to evaluate the performance of our proposed model, we compare it against a set of established methods including domain-specific methods and state-of-the-art MLLMs. 
\begin{itemize}[leftmargin=*]
    \item \textbf{Domain-specific Methods:} We consider four domain-specific methods for ECG including three signal-based methods: METS~\citep{li2024frozen}, MERL~\citep{liu2024zero}, ST-MEM~\citep{na2023guiding}, and one image-based method: ECG-GPT~\citep{khunte2024automated}. 
    \item \textbf{Proprietary MLLMs:} We consider three proprietary MLLMs: GPT-4o, GPT-4o mini~\citep{openai2024gpt4o}, Gemini 1.5 Pro~\citep{reid2024gemini}, 
    and Claude 3.5 Sonnet~\citep{anthropic2024claude}.
    \item \textbf{Open-source MLLMs:} 
    We select a range of open-source models to ensure comprehensive coverage across different model sizes and visual components, including the general models LLaVA-1.5~\citep{liu2024visual,liu2024improved}, LLaVA-1.6~\citep{liu2024llavanext}, Phi-3-Vision~\cite{abdin2024phi}, Idefics2-8B~\citep{laurenccon2024matters}, DeepSeek-Vl-7B~\citep{lu2024deepseekvl}, Mantis-8B-siglip-Llama3~\citep{jiang2024mantis}, MiniCPM-V-2.6~\citep{yao2024minicpm}, InternVL2~\citep{chen2023internvl,chen2024far} and state-of-the-art multimodal models LLaVA-OneVision~\citep{li2024llava}, Qwen2-VL~\citep{Qwen2VL}, as well as the domain-specific models LLaVA-Med~\citep{li2024llava-med}. 
    
    
\end{itemize}

\subsection{Implementation Details}
We follow the architecture of LLaVA-v1.6-Vicuna-7B, which includes three core components: a vision encoder, a large language model, and a projector to align image and text modalities. We format all datasets into a chatbot-style multi-turn dialogue format and use the special token ``$<$image$>$'' to represent image features within the text data. We utilize anyres to support the model's ability to recognize ECG images of various sizes that may appear in real-world scenarios. We freeze the parameters of the vision encoder and fine-tune all parameters of the projector and LLM. We use a learning rate of 2e-5, set the batch size to 128, and employ a cosine scheduler with a 5\% warm-up period for three epochs.

\subsection{Main Results}
We show in-domain the out-of-domain results in Table \ref{tab:in-domain} and Table \ref{tab:out-of-domain} respectively. Overall, we observe that \model achieves state-of-the-art performance on different datasets and tasks. 

\textbf{Results on In-domain datasets.}
As shown in Table \ref{tab:in-domain}, \model demonstrates significant improvements over both proprietary and open-source MLLMs across all in-domain datasets. Specifically, \model surpasses the best proprietary model (GPT-4o) with a 27\% improvement in AUC, an 11-point gain in report score, and a 39\% increase in accuracy on the PTB-XL Super, PTB-XL Report, and ECG-QA tasks, respectively. Moreover, \model achieves notable gains over the best open-source model, with a 28\% improvement in AUC, a 12-point gain in report score, and a 44\% increase in accuracy on the same tasks. 

These results highlight the complexity of ECG image interpretation, a task where even the best proprietary models perform near randomly. By fine-tuning on \data, \model achieves substantial performance improvements, demonstrating the importance of high-quality and task-related instruction tuning. Moreover, while certain domain-specific methods (e.g., MERL) achieve comparable performance on specific datasets, their specialized designs limit their generalization to other diverse tasks, restricting their broader applicability in real-world, complex healthcare scenarios. 

\begin{table*}[!t]
\begin{center}
\adjustbox{max width=\linewidth}{
\begin{tabular}{lccccccccc}
\toprule
Datasets & \multicolumn{3}{c}{PTB-XL Super} & PTB-XL Report & \multicolumn{3}{c}{CODE-15\%} & ECG-QA \\ \midrule
Metric & AUC & F1 & HL & Report Score & AUC & F1 & HL & Accuracy \\ \midrule
Random & 50.3 & 33.2 & 50.1 & 0 & 48.8 & 15.0 & 32.1 & 16.2 \\
\midrule
\multicolumn{8}{c}{Domain-specific Methods} \\ \midrule
METS  & - & 65.7$^\dagger$ & - & N/A & - & - & - & N/A \\
MERL  & 74.2$^\dagger$ & - & - & N/A & - & - & - & N/A \\
ST-MEM  & 71.4$^\dagger$ & - & - & N/A & - & - & - & N/A \\
ECG-GPT & 69.5$^*$ & 53.9$^*$ & 20.1$^*$ & 47.8$^*$ & 68.9$^*$ & 40.1$^*$ & 17.4$^*$ & N/A \\ \midrule
\multicolumn{8}{c}{Proprietary MLLMs} \\ \midrule
GPT-4o & \underline{55.6} & \underline{28.3} & \underline{26.2} & \underline{50.2} & \underline{59.9} & \underline{24.9} & 15.7 & \underline{35.2} \\
GPT-4o mini & 52.0 & 20.4 & 31.7 & 37.1 & 57.5 & 22.0 & \underline{15.1} & 14.9 \\
Gemini 1.5 Pro & 50.7 & 15.3 & 27.9 & 35.9 & 56.7 & 20.0 & 15.9 & 33.2 \\ 
Claude 3.5 Sonnet & 54.0 & 27.5 & 29.6 & 43.7 & 58.3 & 20.3 & 17.8 & 34.2 \\
\midrule
\multicolumn{8}{c}{Open-source MLLMs} \\ \midrule
LLaVA-Med & 50.0 & 12.3 & 28.1 & 24.3 & 69.2 & 27.0 & 33.4 & \underline{29.5} \\
LLaVA-1.5-7B & 50.0 & 12.3 & 28.1 & 27.2 & 63.9 & 19.2 & 25.3 & 25.2 \\
LLaVA-1.5-13B & 50.0 & 35.2 & 48.4 & 20.7 & 53.9 & 13.1 & 13.6 & 21.2 \\
LLaVA-1.6-Vicuna-7B & 50.0 & 15.8 & 29.4 & 16.5 & 50.1 & 1.0 & 13.6 & 13.3 \\
LLaVA-1.6-Vicuna-13B & 50.0 & 20.1 & 38.3 & 5.9 & 53.0 & 3.6 & 16.6 & 22.0 \\
LLaVA-1.6-34B & 50.2 & 19.9 & 36.0 & 17.0 & 57.2 & 12.8 & 16.6 & 22.4 \\
LLaVA-OneVision-7B & 49.8 & 11.4 & 34.5 & 30.0 & 58.7 & 17.0 & 20.6 & 20.4 \\
LLaVA-OneVision-72B & 50.6 & 29.6 & 50.4 & 40.6 & 52.3 & 7.0 & \underline{13.1} & 25.0 \\
Deepseek-VL-Chat-7B & 50.9 & 15.7 & 27.9 & 15.6 & 63.7 & \underline{27.5} & 22.4 & 21.1 \\
Idefics2-8B & 50.7 & 21.9 & 31.2 & 10.6 & 49.0 & 17.9 & 47.9 & 26.1 \\
Mantis-8B-siglip-Llama3 & 50.6 & 20.4 & 30.0 & 16.0 & 57.5 & 17.9 & 15.7 & 23.8 \\
MiniCPM-V-2.6 & 49.0 & \underline{37.7} & 63.8 & 15.4 & 56.6 & 25.3 & 22.0 & 20.8 \\
Phi-3-Vision-128k-Instruct & 50.0 & 29.6 & 48.4 & 20.2 & \underline{69.6} & 22.6 & 38.8 & 28.4 \\
Qwen2-VL-7B & 51.3 & 22.4 & 30.8 & 43.0 & 60.7 & 24.8 & 20.5 & 20.4 \\
Qwen2-VL-72B & \underline{54.0} & 28.3 & 30.2 & \underline{48.9} & 60.6 & 23.6 & 16.1 & 23.7 \\
InternVL2-8B & 50.6 & 14.3 & \underline{27.8} & 38.1 & 55.8 & 16.1 & 17.7 & 22.3 \\
InternVL2-40B & 51.2 & 18.7 & 34.6 & 41.8 & 56.7 & 16.2 & 17.4 & 18.2 \\
InternVL2-Llama3-76B & 50.4 & 9.4 & 35.6 & 41.4 & 59.0 & 20.2 & 20.5 & 21.8 \\ \midrule
\model-7B (Ours) 
& \textbf{82.4} & \textbf{74.8} & \textbf{11.0} & \textbf{61.3} & \textbf{90.7} & \textbf{85.4} & \textbf{5.0} & \textbf{73.8} \\
\rowcolor{blue!10}$\Delta$ over best proprietary MLLM & +27 & +47 & +15 & +11 & +30 & +61 & +10 & +39\\
\rowcolor{blue!10}$\Delta$ over best open-source MLLM& +28 & +37 & +17 & +12 & +21 & +58 & +8 & +44\\
\bottomrule
\end{tabular}}
\end{center}
\caption{In-domain evaluation results. $^\dagger$ indicates results from original papers, $^*$ denotes results obtained using the provided online software, N/A indicates methods not applicable or not designed for certain tasks, and - indicates unreported scores in original papers. Note that the setup of some domain-specific methods is not the same as ours, thus the results listed are for reference purposes. 
}
\label{tab:in-domain}
\end{table*}

\textbf{Results on Out-of-domain datasets.}
Table \ref{tab:out-of-domain} presents the comparison results on out-of-domain datasets, where \model consistently delivers outstanding performance. Notably, it achieves a significant 15\% improvement in accuracy on the MMMU ECG benchmark compared to GPT-4o. This substantial improvement indicates the \model's robustness and ability to generalize to unseen data. 

The ECG Arena benchmark presents a significantly more challenging task for all models. This benchmark is characterized by its multi-turn, open-ended question-answering format, which closely simulates real clinical scenarios. Despite these challenges, \model still surpasses the best proprietary model by 2 points and outperforms the leading open-source model by an impressive 11 points in terms of arena score. These results highlight \model's relative strength in handling complex, clinically-oriented ECG interpretation and analysis. Additionally, the performance gap across models on this challenging benchmark indicates considerable room for future improvements in this task.

\begin{table*}[!t]
\begin{center}
\adjustbox{max width=\linewidth}{
\begin{tabular}{@{}lccccccc@{}}
\toprule
Datasets & \multicolumn{3}{c}{CPSC 2018} & CSN & G12EC & \multicolumn{1}{l}{MMMU ECG} & ECG Arena \\ \midrule
Metric & AUC & F1 & HL & Accuracy & Accuracy & Accuracy & Arena Score \\ 
\midrule
Random & 51.2 & 15.1 & 28.8 & 11.6 & 12.1 & 24.2 & 0 \\
\midrule
\multicolumn{8}{c}{Domain-specific Methods} \\ \midrule
METS & - & - & - & N/A & N/A & N/A & N/A \\
MERL & 82.8$^\dagger$ & - & - & N/A & N/A & N/A & N/A \\
ST-MEM  & 70.4$^\dagger$ & - & - & N/A & N/A & N/A & N/A \\
ECG-GPT & 69.3$^*$ & 44.0$^*$ & 9.9$^*$ & N/A & N/A & N/A & N/A \\ \midrule
\multicolumn{8}{c}{Proprietary MLLMs} \\ \midrule
GPT-4o & 50.9 & 10.6 & \underline{18.2} & \underline{57.5} & 49.2 & \underline{43.5} & 33.5 \\
GPT-4o mini & 49.2 & 11.0 & 25.5 & 32.1 & 33.2 & 39.5 & 30.1 \\
Gemini-1.5-Pro & 50.1 & 7.4 & 20.5 & 50.5 & 36.0 & 40.0 & 31.2 \\ 
Claude 3.5 Sonnet & \underline{52.8} & \underline{11.5} & 18.9 & 51.5 & \underline{51.4} & 42.0 & \underline{37.1} \\ \midrule
\multicolumn{8}{c}{Open-source MLLMs} \\ \midrule
LLaVA-Med & 50.0 & 2.5 & 20.2 & 13.8 & 14.1 & 27.0 & 15.9 \\
LLaVA-1.5-7B & 50.0 & 2.5 & 20.0 & 32.1 & 25.4 & 33.0 & 12.7 \\
LLaVA-1.5-13B & 50.4 & 13.3 & 30.1 & 30.7 & 30.7 & 35.0 & 13.1 \\
LLaVA-1.6-Vicuna-7B & 50.5 & \underline{19.7} & 66.0 & 23.7 & 23.3 & 28.0 & 16.0 \\
LLaVA-1.6-Vicuna-13B & 50.0 & 19.3 & 62.8 & 31.4 & 35.0 & 38.0 & 17.9 \\
LLaVA-1.6-34B & 49.6 & 19.3 & 62.8 & 44.3 & \underline{45.9} & 31.0 & 17.5 \\
LLaVA-OneVision-7B & 49.6 & 8.0 & 28.3 & 23.3  & 25.7 & 26.0 & 22.5 \\
LLaVA-OneVision-72B & 51.5 & 12.8 & 29.4 & 44.0  & 42.6 & 35.0 & 15.5 \\
Deepseek-VL-Chat-7B & 50.7 & 6.0 & 20.0 & 35.7 & 32.9 & 34.5 & 15.3 \\
Idefics2-8B & 49.0 & 17.9 & 47.9 & 22.8 & 26.2 & 36.0 & 4.9 \\
Mantis-8B-siglip-Llama3 & 51.3 & 19.1 & 48.5 & 17.6 & 22.6 & \underline{38.5} & 13.6 \\
MiniCPM-2.6 & 50.0 & 18.0 & 48.4 & 12.7 & 19.6 & 34.5 & 20.4 \\
Phi-3-Vision-128k-Instruct & 50.6 & 19.0 & 70.2 & 14.8 & 18.4 & 31.0 & 11.3 \\
Qwen2-VL-7B & 49.4 & 17.5 & 46.3 & 25.5 & 32.9 & 31.5 & 8.5 \\
Qwen2-VL-72B & 50.7 & 9.8 & \underline{18.9} & 35.5 & 42.9 & 35.0 & 10.3 \\
InternVL2-8B & 52.1 & 8.2 & 22.2 & \underline{47.7}  & 37.5 & 30.0 & 22.9 \\
InternVL2-40B & \underline{52.4} & 8.2 & 21.4 & 41.0 & 45.0 & 30.5 & \underline{28.0} \\
InternVL2-Llama3-76B & 51.3 & 6.5 & 20.4 & 26.6 & 34.7 & 38.0 & 22.5 \\ \midrule
\model-7B (Ours) & \textbf{76.9} & \textbf{57.6} & \textbf{8.6} & \textbf{85.2} & \textbf{78.2} & \textbf{58.0} & \textbf{38.9}\\ 
\rowcolor{blue!10}$\Delta$ over best proprietary MLLM & +24 & +46 & +10 & +28 & +27 & +15 & +2 \\
\rowcolor{blue!10}$\Delta$ over best open-source MLLM & +25 & +38 & +10 & +38 & +33 & +20 & +11 \\
\bottomrule
\end{tabular}
}
\end{center}
\caption{Out-of-domain evaluation results. $^\dagger$ indicates results from original papers, $^*$ denotes results obtained using the provided online software, N/A indicates methods not applicable or not designed for certain tasks, and - indicates unreported scores in original papers.}
\label{tab:out-of-domain}
\end{table*}

\subsection{Ablation Study}

\textbf{Effect of Training Data Source.}
Given that \data is compiled from diverse datasets, it is crucial to examine how each dataset contributes to the model's overall performance. Table \ref{tab:ablation_train_data} presents a comparative analysis of models trained on various dataset combinations. The model trained exclusively on PTB-XL (P) exhibits the lowest performance across all datasets, indicating the limitations of relying on a single data source for effective generalization.
As we progressively incorporate additional datasets into the training set, the model's performance consistently improves. These results highlight the importance of curating diverse training data, as expanding beyond a single source enhances the model’s capacity to generalize across datasets and tasks.


\begin{table*}[!t]
    \begin{center}
    \adjustbox{max width=\linewidth}{
    \begin{tabular}{lccccccccc|c}
        \toprule
        Training Data & \begin{tabular}[c]{@{}c@{}}PTB-XL\\ Super\end{tabular} & \begin{tabular}[c]{@{}c@{}}PTB-XL\\ Report\end{tabular} & CSN & CODE-15 & ECQ-QA & CPSC & G12 & \begin{tabular}[c]{@{}c@{}}MMMU \\ ECG\end{tabular} & \begin{tabular}[c]{@{}c@{}}ECG \\ Arena\end{tabular} & \textbf{AVG} \\
        \midrule
        P & 68.2 & 56.7 & 82.8 & 31.5 & 31.8 & 23.4 & 65.4 & 40.0 & 28.4 & \textbf{-20.6}\\
        P + M & 74.1 & 62.4 & 88.7 & 48.5 & 35.8 & 52.4 & 78.8 & 58.5 & 37.0 & \textbf{-8.6}\\
        P + M + C & 74.1 & 63.8 & 87.5 & 85.8 & 43.4 & 51.0 & 75.5 & 55.5 & 39.4 & \textbf{-4.1} \\ \midrule
        P + M + C + E & 74.8 & 61.3 & 85.2 & 85.4 & 73.8 & 57.6 & 78.2 & 58.0 & 38.9 & \textbf{68.1} \\
        \bottomrule
    \end{tabular}}
    \end{center}
    \caption{Performance of different training dataset combinations. P: PTB-XL, M: MIMIC-IV-ECG, C: CODE-15\%, E: ECG-QA. F1 for PTB-XL Super, CODE-15\%, and CPSC; Accuracy for CSN, ECG-QA, G12, and MMMU ECG; Report Scores for PTB-XL Report; Arena Scores for ECG Arena. \textbf{AVG} denotes the average across all metrics.}
    \label{tab:ablation_train_data}
    \vspace{-10pt}
\end{table*}

\textbf{Effect of Instruction Task.}
To understand the individual contribution of each ECG-related task to model performance, we analyze combinations of four instruction tasks. As shown in Table \ref{tab:ablation_task}, adding more tasks progressively improves performance across multiple benchmarks. Models trained solely on basic feature recognition (F) performed poorly across all metrics, highlighting the limitations of a single-task approach. In contrast, the sequential addition of tasks led to substantial performance gains across multiple benchmarks. The model incorporating all four tasks achieved the highest performance, indicating a more comprehensive understanding of ECG images. 

\begin{table*}[!t]
    \begin{center}
    \adjustbox{max width=\linewidth}{
    \begin{tabular}{lccccccccc|c}
        \toprule
        Instruction Task & \begin{tabular}[c]{@{}c@{}}PTB-XL\\ Super\end{tabular} & \begin{tabular}[c]{@{}c@{}}PTB-XL\\ Report\end{tabular} & CSN & CODE-15 & ECQ-QA & CPSC & G12 & \begin{tabular}[c]{@{}c@{}}MMMU \\ ECG\end{tabular} & \begin{tabular}[c]{@{}c@{}}ECG \\ Arena\end{tabular} & \textbf{AVG} \\
        \midrule
        F & 12.3 & 36.0 & 56.6 & 11.2 & 54.8 & 2.5 & 11.2 & 34.0 & 12.4 & \textbf{-42.5}\\
        F + R & 26.9 & 54.0 & 83.8 & 73.3 & 61.4 & 31.0 & 67.3 & 47.5 & 25.3 & \textbf{-15.9}\\
        F + R + M & 70.4 & 57.6 & 85.2 & 82.7 & 68.6 & 43.8 & 71.0 & 52.5 & 30.4 & \textbf{-5.7} \\\midrule
        F + R + M + C & 74.8 & 61.3 & 85.2 & 85.4 & 73.8 & 57.6 & 78.2 & 58.0 & 38.9 & \textbf{68.1} \\
        \bottomrule
    \end{tabular}}
    \end{center}
    \caption{Performance of different ECG-related instruction task combinations. F: basic feature recognition, R: heart rhythm analysis, M: morphology and pathology identification, C: clinical report generation. F1 for PTB-XL Super, CODE-15\%, and CPSC; Accuracy for CSN, ECG-QA, G12, and MMMU ECG; Report Scores for PTB-XL Report; Arena Scores for ECG Arena. \textbf{AVG} denotes the average across all metrics.}
    \label{tab:ablation_task}
    \vspace{-10pt}
\end{table*}


\subsection{Case Study}
We further present some examples from our benchmark, comparing the outputs of our model with GPT-4o for ECG report generation (Appendix Figs. \ref{fig:test_example_report_10}-\ref{fig:test_example_report_5338}) and ECG Arena (Appendix Fig. \ref{fig:test_example_arena_1}). While GPT-4o is capable of generating reports and answering questions by following instructions, it often produces responses that, although well-structured and seemingly relevant, contain significant inaccuracies in interpretation. In contrast, \model consistently provides more accurate responses that align closely with the ground truths. Additionally, we observed that GPT-4o tends to over-rely on its OCR capabilities when textual information (e.g., printed axis labels, numerical values like heart rate or QRS duration) is present in images, leading to superficial reasoning based on text rather than a deep analysis of visual data. As shown in Appendix Fig. \ref{fig:test_example_report_5338}, GPT-4o identifies a left axis deviation based on the printed QRS axis degree, without analyzing the visual waveform patterns. If such axis information were absent, the model would likely fail to identify the deviation.

\subsection{Discussion}
\begin{wrapfigure}{r}{0.4\textwidth}
  \centering
  \includegraphics[width=0.38\textwidth]{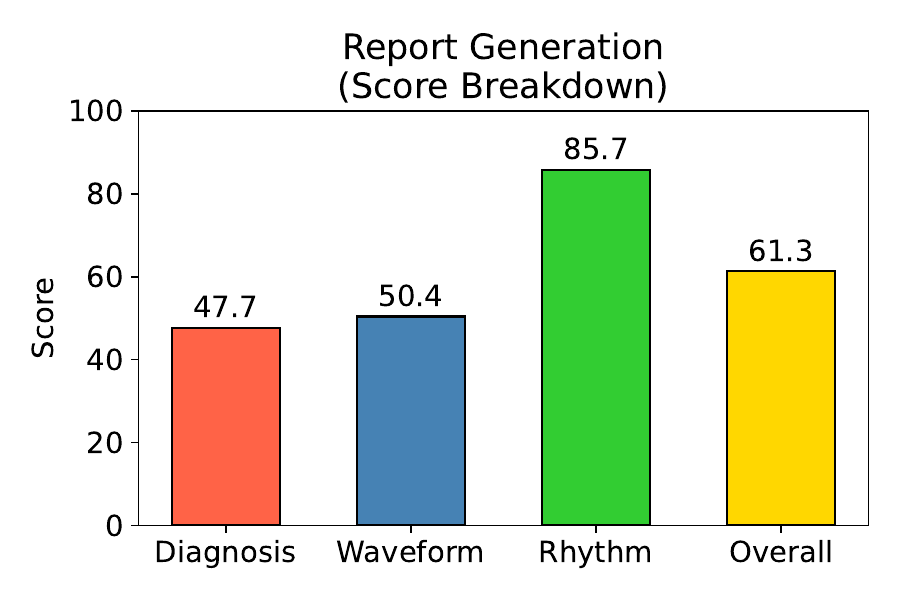}
  \caption{Score breakdown of report generation performance.}
\label{fig:report_score_breaksown}
\end{wrapfigure}
While the model demonstrates superior performance across various evaluation datasets, it faces notable challenges with more complex and open-ended tasks, such as report generation and multi-turn conversations. To further investigate the model's performance in report generation, we present the score breakdown in Fig. \ref{fig:report_score_breaksown}. The model excels in rhythm interpretation but struggles with waveform and diagnosis identification. These results suggest that future efforts should prioritize increasing the dataset's coverage of waveform and diagnosis-related cases to enhance the model's ability to detect these abnormalities. Additionally, as diagnosis identification may require more advanced multi-step reasoning, future research could focus on incorporating step-wise instruction tuning data to strengthen the model's reasoning capabilities. 

\section{Conclusion}
In this paper, we study the problem of ECG image interpretation, which is a crucial task in assessing cardiac conditions. We develop a new MLLM, \model, fine-tuned on the newly created \data dataset with over 1 million samples across a diverse range of ECG-related tasks. Evaluated on the proposed benchmark, \eval, our model shows state-of-the-art performance, surpassing both proprietary and open-source MLLMs across multiple in-domain and out-of-domain evaluation datasets. This work demonstrates the potential of using MLLMs for enhancing ECG image analysis and interpretation in clinical applications.

\bibliographystyle{iclr2025_conference}

\newpage
\appendix

\renewcommand{\thefigure}{A\arabic{figure}}
\setcounter{figure}{0}
\renewcommand{\thetable}{A\arabic{table}}
\setcounter{table}{0}
\setcounter{equation}{0}
\setcounter{footnote}{0}

\DoToC

\section{Related Work}
\paragraph{Domain-specific Models for ECG.} 

Many domain-specific models have been proposed to enhance automatic ECG diagnosis~\citep{hannun2019cardiologist, ribeiro2020automatic, hughes2021performance}. For example, \citet{ribeiro2020automatic} applied convolutional neural networks (CNNs) to encode ECG signals for diagnosing 6 types of abnormalities. To reduce dependence on
high-quality labeled data, recent studies~\citep{li2024frozen, liu2024zero, na2023guiding} have further explored self-supervised learning approaches using unlabeled ECG training data. For example, \citet{liu2024zero} proposed an ECG representation learning framework by integrating the ECG signals and clinical reports, showing improved performance in zero-shot ECG classification tasks. Despite these successes, most approaches treat ECG data as temporal physiological signals, which could be limiting in certain resource-constrained or remote settings where only printed or digital images are available. Recently, a few methods~\citep{sangha2022automated, sangha2023detection, khunte2024automated} have been proposed for ECG diagnosis using ECG images. For example, \citet{khunte2024automated} developed a diagnostic report generation framework for ECG images, which is built upon a BEiT~\citep{bao2021beit} vision transformer encoder and a GPT-2~\citep{radford2019language} decoder. However, their model is only capable of the clinical report generation task, without generalizability to other diverse tasks. In contrast, our study investigates the capabilities of MLLMs for ECG image interpretation. We curate a diverse range of instruction tuning datasets to fine-tune the model, thus improving model generalizability.

\paragraph{MLLMs in Healthcare}
Recent advancements in MLLMs have shown promising results in various healthcare domains. General medical multimodal models such as LLaVA-Med~\citep{li2024llava}, MedPaLM~\citep{singhal2023large, singhal2023towards}, and Med-Gemini~\citep{saab2024capabilities} have demonstrated capabilities in processing diverse medical data types. Additionally, domain-specific multimodal models have been developed for specialized fields like pathology~\citep{lu2024multimodal, xu2024whole} and radiology~\citep{wu2023towards}. These models have shown potential in integrating visual and textual information to support clinical decision-making and medical education. However, despite the importance of ECG data in cardiac diagnosis and monitoring, current MLLMs often struggle to process ECG images effectively. This limitation highlights a significant gap in the application of MLLMs to cardiology, where the ability to interpret both visual ECG representations and accompanying clinical information is crucial. 

\paragraph{Multimodal Instruction Tuning.}
Instruction tuning has proven effective in the multimodal domain, particularly in vision-language models like LLaVA~\citep{liu2024visual}, MiniGPT-4~\citep{zhu2023minigpt} and InstructBLIP~\citep{dai2023instructblip}. These models demonstrate impressive generalizability across various visual understanding and reasoning tasks. While multimodal instruction tuning has been applied to general medical imaging tasks~\citep{li2024llava-med, singhal2023large}, its application to ECG images remains largely unexplored. A recent work~\citep{wan2024electrocardiogram} introduced a targeted instruction tuning framework and fine-tuned existing open-source LLMs for ECG report generation. However, their approach is limited by a single-task instruction dataset focused solely on report generation, potentially constraining its adaptability to other ECG-related tasks. Moreover, their work also treats ECG data as temporal signals, whereas our paper focuses on encoding ECG images with MLLMs, which is more applicable to real scenarios where only printed or digital ECG images are available.

\section{Preliminary on 12-lead ECG}
ECG is a vital diagnostic tool that measures the electrical activity of the heart over time, providing insights into both spatial and temporal aspects of cardiac function. Typically, an ECG recording is presented as a 12-lead multivariate time series, where each lead offers a unique perspective on heart activity. The six limb leads (I, II, III, aVR, aVL, and aVF) assess the electrical movements across the arms and legs, giving views from the frontal plane. Simultaneously, the six precordial leads (V1, V2, V3, V4, V5, and V6) monitor the chest, offering horizontal plane views. In this paper, we focus on ECG images that are synthesized from raw signals. 

\section{Details of ECG Image Synthesis}
\label{sec:image_synthesis}

We employ the ECG-image-kit~\citep{shivashankara2024ecg} framework to synthesize diverse ECG images from raw signal data. This toolkit allows for the generation of ECG images under various conditions by introducing a range of distortions and noises to better simulate real-world clinical data. 

Specifically, in addition to generating standard 12-lead ECG images—characterized by black waveforms on a white background, red grid lines, and a 4x3 layout—we introduce a variety of perturbations to the images. These modifications include the addition of wrinkles and creases, simulating the physical wear and tear commonly observed in paper-printed ECGs. 
Our image synthesis process includes various augmentation methods to simulate physical distortions, image quality variations, and layout alterations. We introduce wrinkles and creases to mimic wear and tear commonly observed in paper-printed ECGs, and apply random rotations at varying angles to simulate misaligned scans or prints. To account for different acquisition systems and scanning qualities, we vary image resolutions and introduce random background colors, such as slight yellowing to represent aging or poor scanning quality. We also add noise to the images to simulate imperfections in the scanning or printing process. Furthermore, we experiment with different aspect ratios, overall image sizes, and ECG plot positions within the image to reflect the heterogeneity of ECG printouts across different systems and formats. In some cases (with a 0.02 probability), we randomly remove grid lines to account for variations in ECG presentation. 

To further enrich the synthetic images, we randomly insert meta-information into the image header to simulate the annotations typically seen in clinical ECG reports. For the PTB-XL dataset, we extract patient demographics (e.g., age, gender) and basic ECG features (e.g., heart rate, axis deviations) from the associated PTB-XL feature annotation dataset, PTB-XL+~\citep{strodthoff2023ptb}. This extracted data is used to impute realistic meta-information, which is then randomly printed on the synthesized image. This random insertion of meta-data not only increases the visual variety of the images but also provides additional context, simulating real-world ECG prints that include patient and diagnostic information. To further increase diversity, we adopt alternative lead configurations beyond the standard 4x3 layout, such as 12x1 (single row of leads), 6x2 (two rows of six leads), and other commonly used clinical formats. These variations ensure that our model is exposed to a wide range of ECG presentation styles.

The augmentation process is designed to balance the dataset, with an approximate ratio of 1:1 between augmented and standard ECG images. This balance ensures that the model is exposed to both clean and distorted images, aiding in its generalization to real-world clinical scenarios.

\clearpage
\section{Details of Instruction Tuning Datasets}

\begin{table*}[!th]
\begin{center}
\adjustbox{max width=\linewidth}{
\begin{tabular}{@{}llll@{}}
\toprule
Source   Dataset & Task & Type & \# Samples \\ \midrule
\multirow{13}{*}{PTB-XL} & Basic Feature Recognition & Close-ended QA & 22,759 \\
 &  & Open-ended QA & 906 \\
 &  & Fill-in-blank & 449 \\
 &  & Multi-choice QA & 5,716 \\ \cmidrule(l){2-4} 
 & Heart Rhythm Analysis & Close-ended QA & 19,550 \\
 &  & Open-ended QA & 201 \\
 &  & Fill-in-blank & 436 \\
 &  & Multi-choice QA & 16,179 \\ \cmidrule(l){2-4} 
 & Morphology and Pathology Identification & Close-ended QA & 50,166 \\
 &  & Open-ended QA & 2,649 \\
 &  & Fill-in-blank & 680 \\
 &  & Multi-choice QA & 13,432 \\ \cmidrule(l){2-4} 
 & Clinical Report & Open-ended QA & 16,302 \\ \midrule
PTB-XL Total &  &  & 149,425 \\ \midrule
\multirow{3}{*}{ECG-QA} & Basic Feature Recognition & Close-ended QA & 40,154 \\
 & Heart Rhythm Analysis & Close-ended QA & 8,993 \\
 & Morphology and Pathology Identification & Close-ended QA & 90,211 \\ \midrule
ECG-QA Total &  &  & 139,358 \\ \midrule
\multirow{13}{*}{MIMIC-ECG} & Basic Feature Recognition & Close-ended QA & 759 \\
 &  & Open-ended QA & 4,759 \\
 &  & Fill-in-blank & 6,470 \\
 &  & Multi-choice QA & 17,186 \\ \cmidrule(l){2-4} 
 & Heart Rhythm Analysis & Close-ended QA & 48,625 \\
 &  & Open-ended QA & 5,262 \\
 &  & Fill-in-blank & 11,487 \\
 &  & Multi-choice QA & 49,352 \\ \cmidrule(l){2-4} 
 & Morphology and Pathology Identification & Close-ended QA & 8,241 \\
 &  & Open-ended QA & 81,080 \\
 &  & Fill-in-blank & 18,264 \\
 &  & Multi-choice QA & 61,456 \\ \cmidrule(l){2-4} 
 & Clinical & Open-ended QA & 486,746 \\ \midrule
MIMIC-ECG Total &  &  & 799,687 \\ \midrule
\multirow{3}{*}{CODE-15\%} & Basic Feature Recognition & Close-ended QA & 22,177 \\
 & Heart Rhythm Analysis & Close-ended QA & 13,893 \\
 & Morphology and Pathology Identification & Close-ended QA & 31,570 \\ \midrule
CODE-15\% Total &  &  & 67,640 \\ \midrule
\data Total &  &  & 1,156,110 \\ \bottomrule
\end{tabular}}
\end{center}
\caption{Detailed data statistics of \data.}
\label{tab:ecg-instruct-stat-all}
\end{table*}

\begin{figure*}[!t]
    \centering
\includegraphics[width=0.99\textwidth]{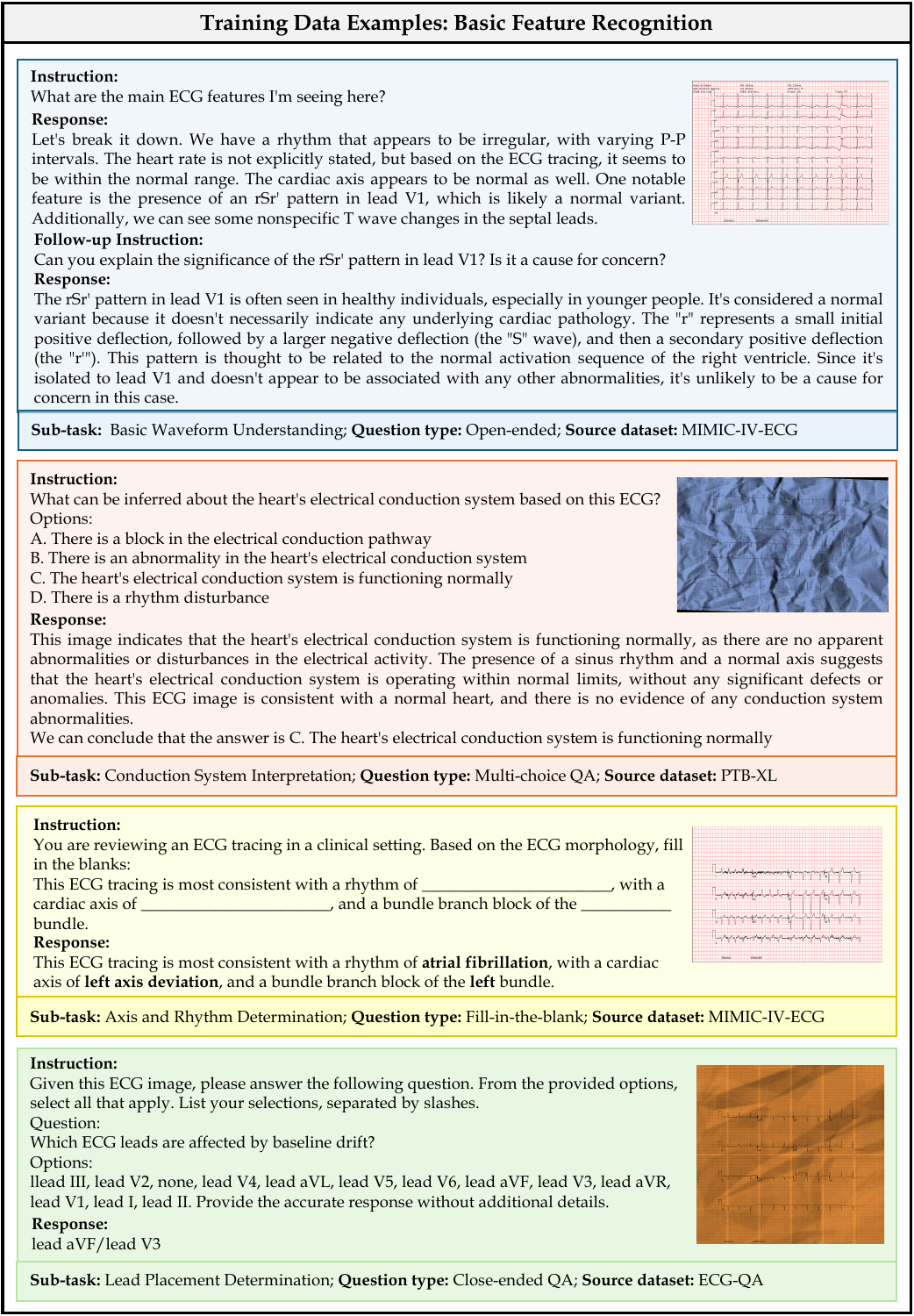}
    \caption{The Examples of basic feature recognition instructions for finetuning \model.}
\label{fig:train_data_basic}
\end{figure*}

\begin{figure*}[!t]
    \centering
\includegraphics[width=0.99\textwidth]{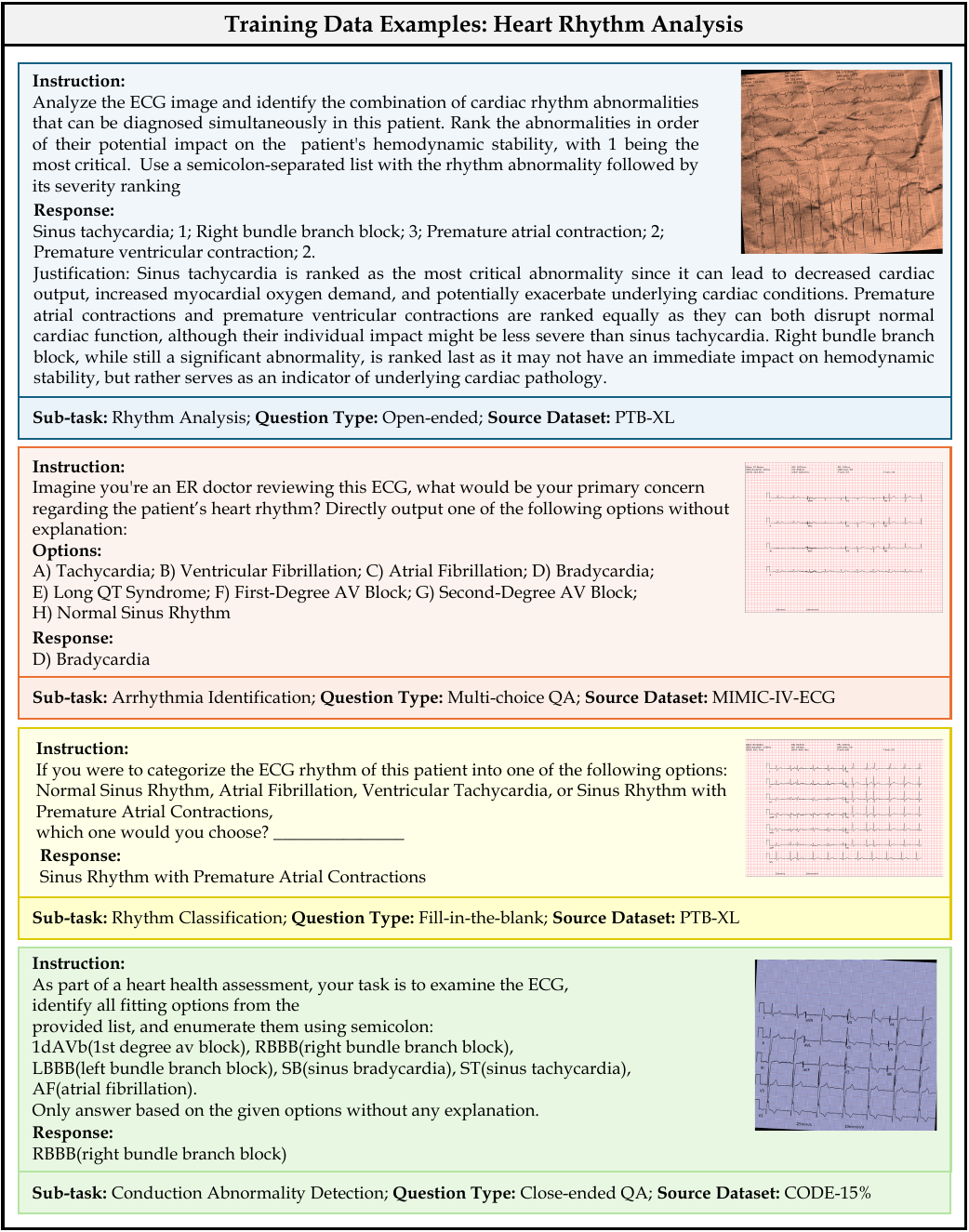}
    \caption{The Examples of heart rhythm analysis instructions for finetuning \model.}
\label{fig:train_data_rhythm}
\end{figure*}

\begin{figure*}[!t]
    \centering
\includegraphics[width=0.99\textwidth]{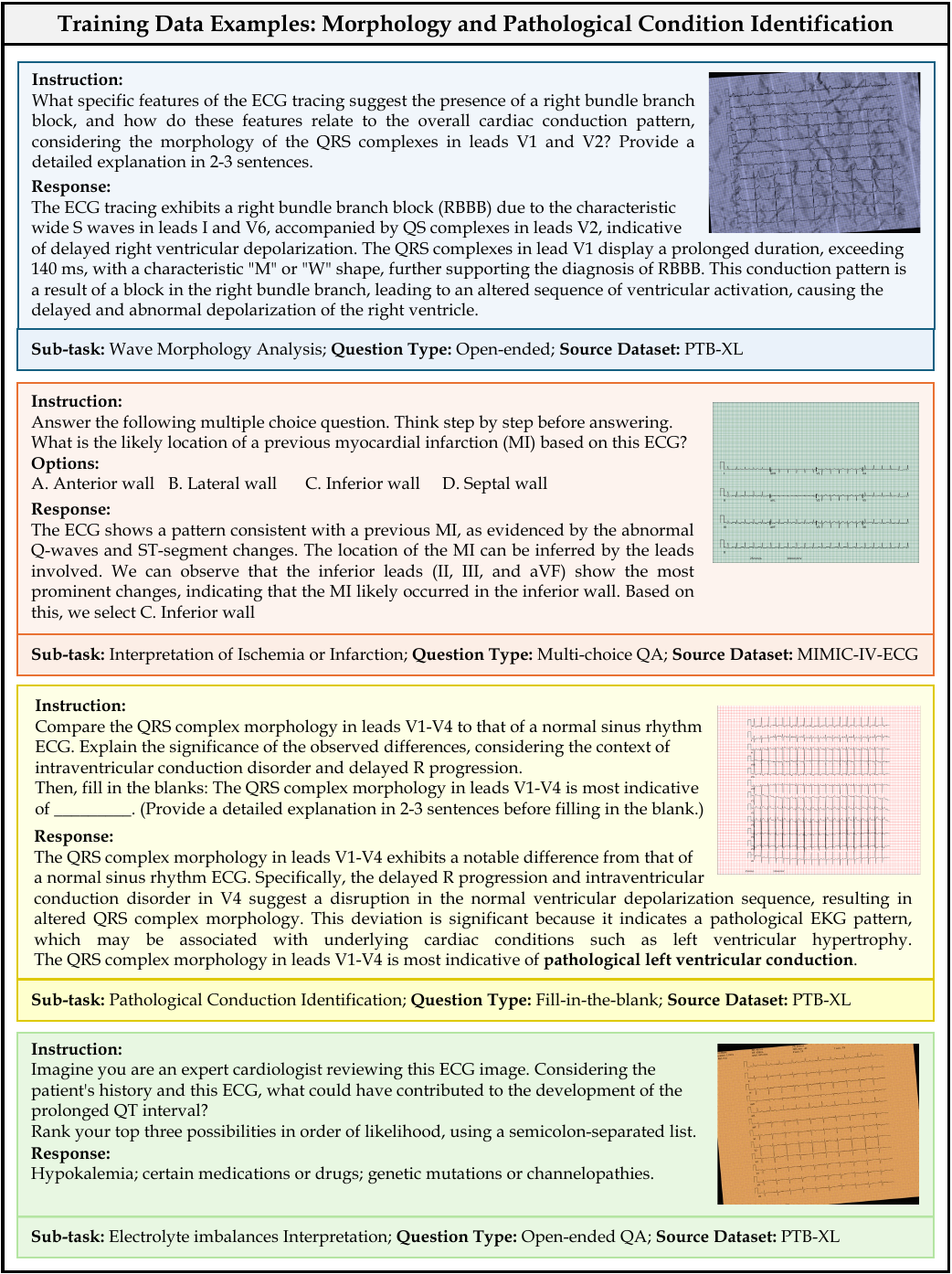}
    \caption{The Examples of morphology and pathological condition identification instructions for finetuning \model.}
\label{fig:train_data_morph_path}
\end{figure*}

\begin{figure*}[!t]
    \centering
\includegraphics[width=0.99\textwidth]{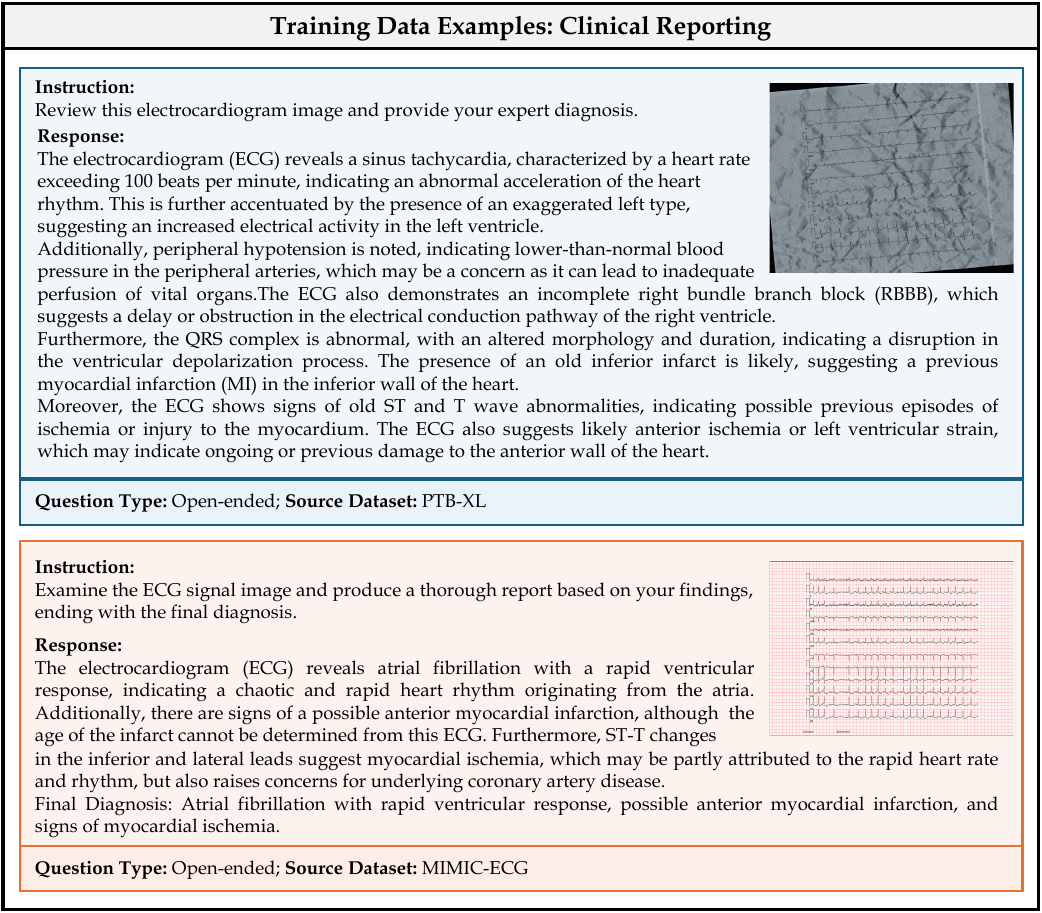}
    \caption{The Examples of clinical reporting instructions for finetuning \model.}
\label{fig:train_data_report}
\end{figure*}

\clearpage

\section{Prompts} \label{sec:prompt}
\begin{figure*}[!ht]
    \centering
    \includegraphics[width=0.99\textwidth]{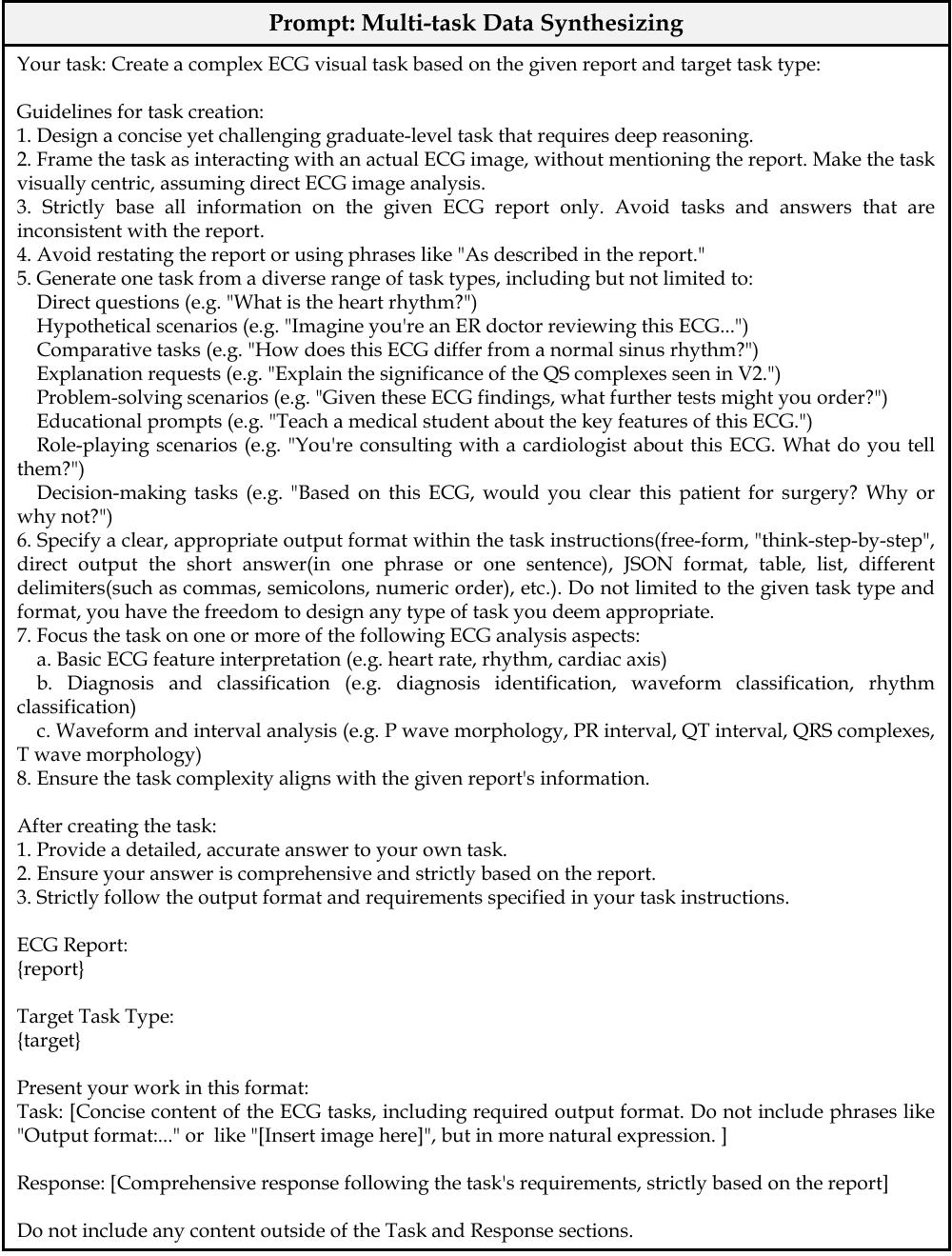}
    \caption{The prompt used to synthesize ECG instruction tasks based on clinical reports.}
    \label{fig:prompt_multi_task_data_syn}
\end{figure*}

\begin{figure*}[!t]
    \centering
    \includegraphics[width=0.99\textwidth]{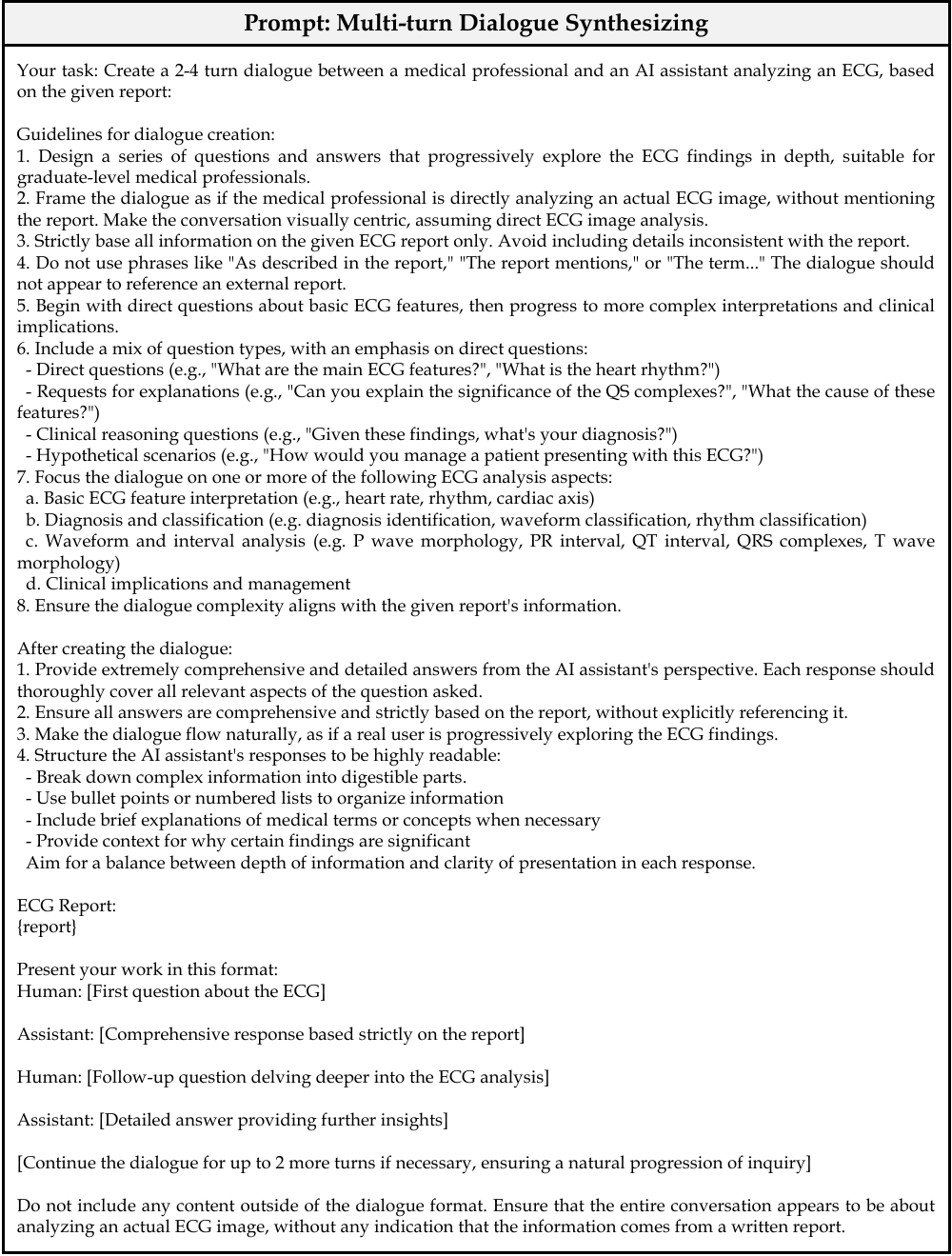}
    \caption{The prompt used to synthesize ECG multi-turn dialogue as instruction tuning data.}
    \label{fig:prompt_multi_turn_diag_syn}
\end{figure*}

\begin{figure*}[!t]
    \centering
    \includegraphics[width=0.99\textwidth]{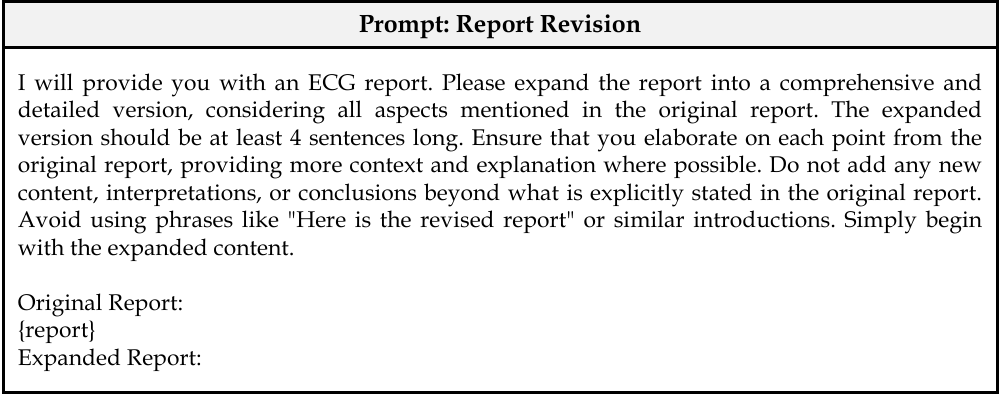}
    \caption{The prompt used to revise (and translate) original reports.}
    \label{fig:prompt_report_revision}
\end{figure*}

\begin{figure*}[!t]
    \centering
    \includegraphics[width=0.99\textwidth]{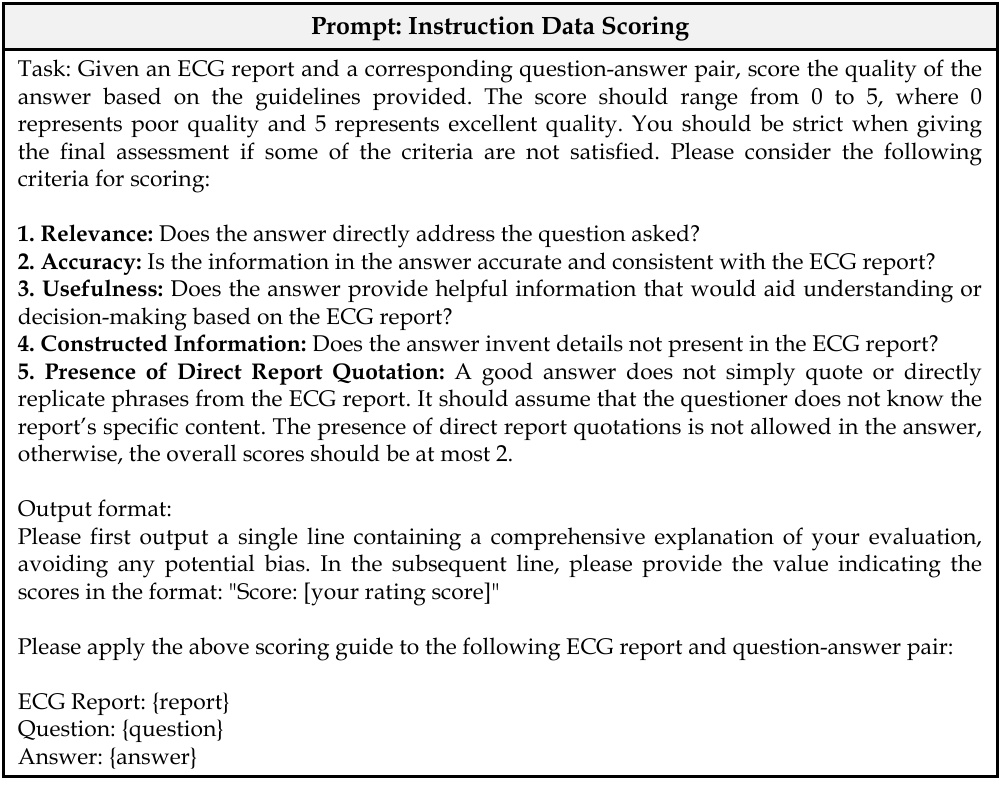}
    \caption{The prompt used to score and filter generated instruction data.}
    \label{fig:prompt_instruction_score}
\end{figure*}

\begin{figure*}[!t]
    \centering
    \includegraphics[width=0.99\textwidth]{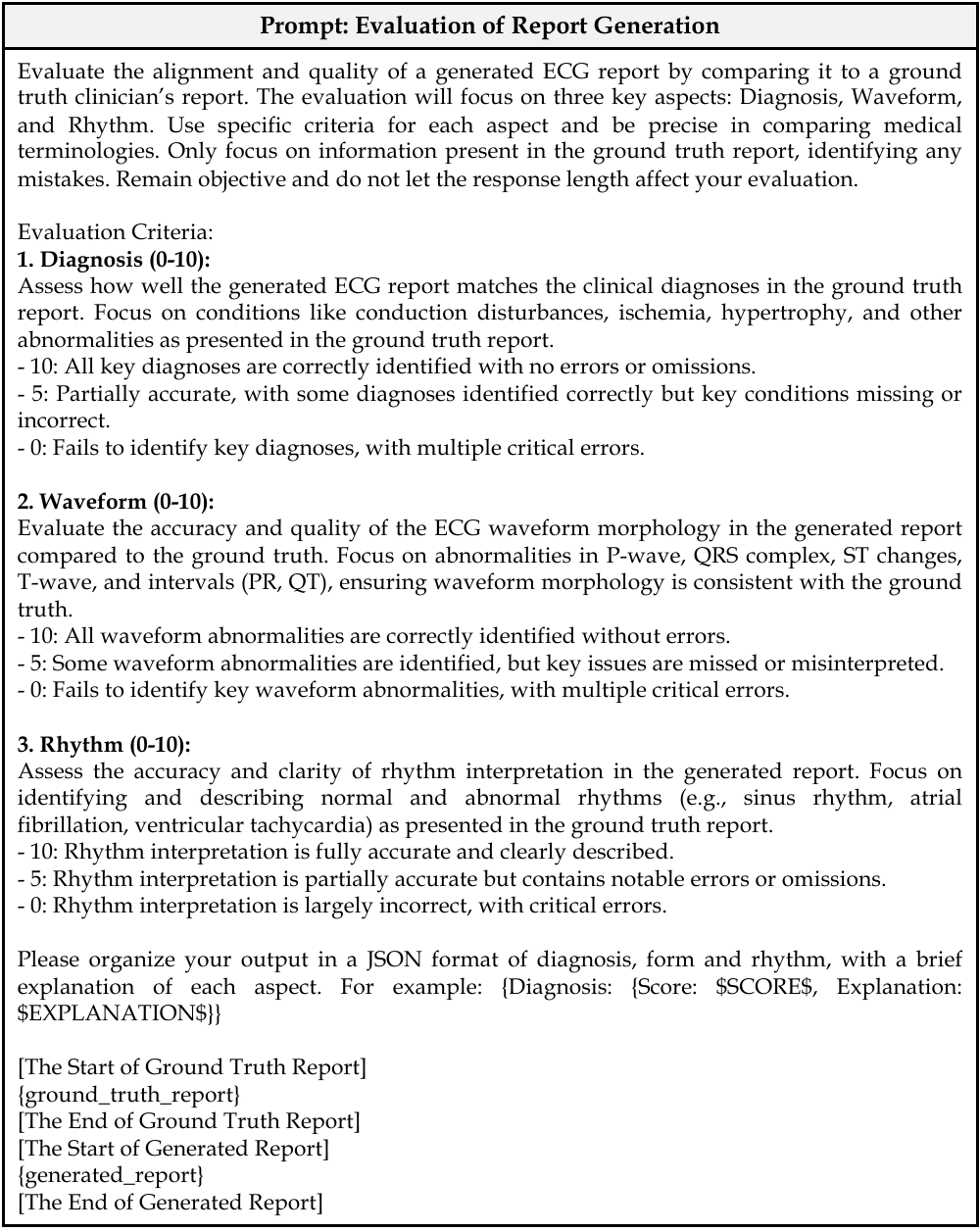}
    \caption{The prompt used to evaluate the generated report.}
    \label{fig:prompt_report_eval}
\end{figure*}

\begin{figure*}[!t]
    \centering
    \includegraphics[width=0.99\textwidth]{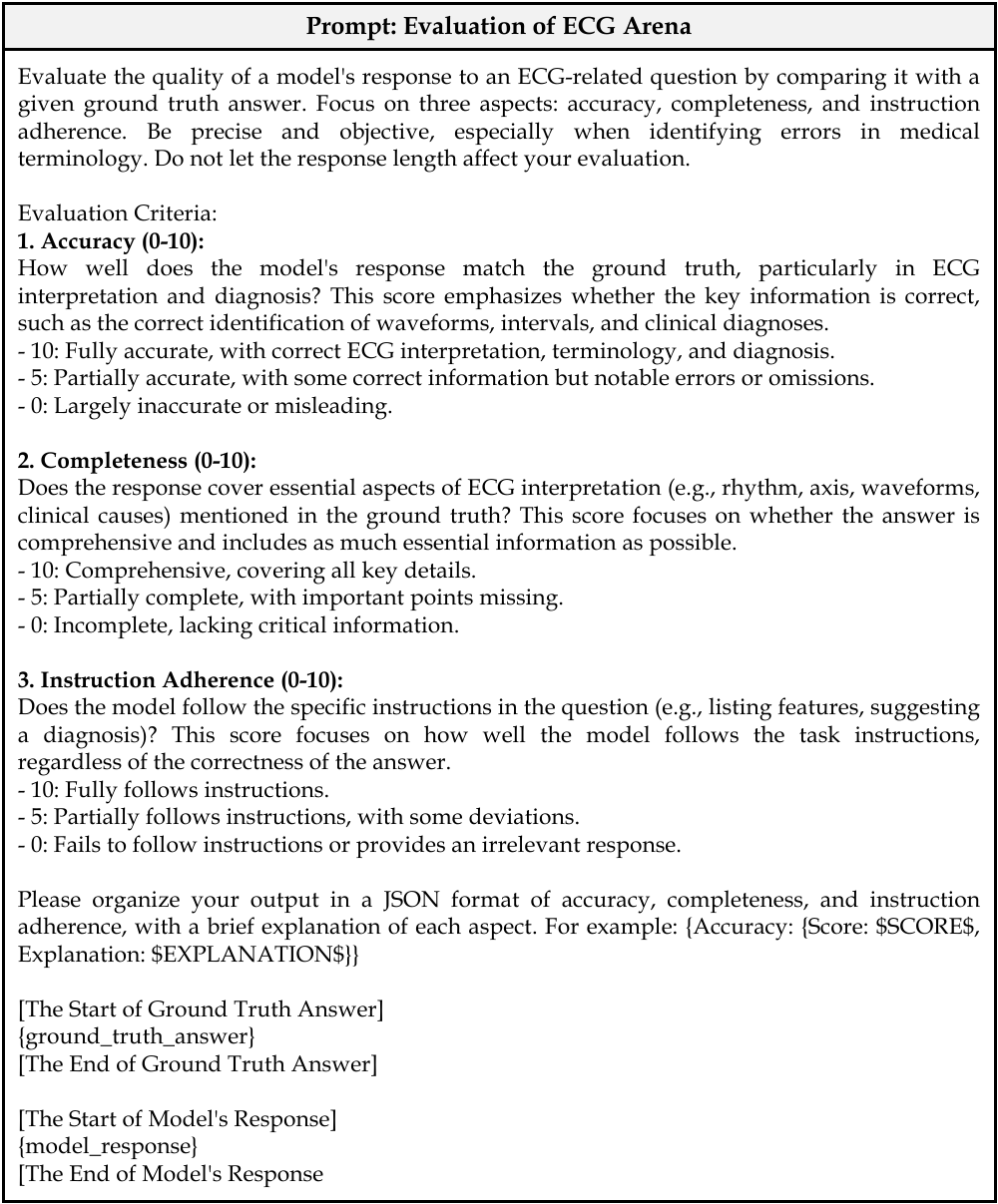}
    \caption{The prompt used to evaluate the ECG Arena.}
    \label{fig:prompt_arena_eval}
\end{figure*}

\clearpage
\section{Details of Evaluation Metrics}
\label{sup:eval_details}
\textbf{Abnormality Detection.} we utilize multi-label classification metrics, including Macro AUC, Macro F1, and Hamming Loss, to evaluate the datasets PTB-XL Super, CODE-15\%, and CPSC 2018, where multiple correct labels may exist. For the ECG-QA, CSN, and G12EC datasets, we adopt accuracy as the evaluation metric.

\textbf{Report Generation.} Rather than relying on traditional text generation metrics, we leverage strong LLMs as evaluators, following the approach of \citet{zheng2024judging}. This method provides a more nuanced evaluation by focusing on key aspects of the reports. Specifically, we use GPT-4o to compare the model-generated reports against those written by cardiologists. We introduce a ``Report Perfect Score'', which is based on three critical components of a generated report: (1) Rhythms (0 to 10 points), (2) Waveform Morphology (0 to 10 points), and (3) Diagnosis (0 to 10 points). The final score is the average of these three components, scaled to a maximum of 100 points. The prompt used to query GPT-4o for evaluating the report score is provided in Appendix Fig. \ref{fig:prompt_report_eval}.

\textbf{MMMU ECG.} We adopt accuracy as the primary metric. We have designed systematic, rule-based evaluation pipelines to ensure robust and consistent scoring. To mitigate the potential influence of any intermediate generations (e.g., reasoning steps) in long responses, we employ robust regular expressions and develop response-processing workflows. These are used to extract answer options from the long responses for accurate answer matching. In cases where no valid answer can be extracted from the model's response, we perform random selection to assign a score. 

\textbf{ECG Arena.} We also employ a strong judge model, GPT-4o, to assess model performance by comparing generated responses with ground truth answers. The evaluation considers three perspectives, each scored on a scale of 0-10: Accuracy (how closely the model's response matches the ground truth), Completeness (whether the model provides a comprehensive answer covering all aspects of ECG interpretation), and Instruction Adherence (how well the model follows the specific instructions in the question). We calculate the final score by averaging these three aspects and scaling to a maximum of 100 points. The specific prompt used for GPT-4 evaluation is provided in Appendix Fig. \ref{fig:prompt_arena_eval}.

\newpage
\clearpage

\section{Case Study}

\begin{figure*}[!th]
    \centering
\includegraphics[width=0.9\textwidth]{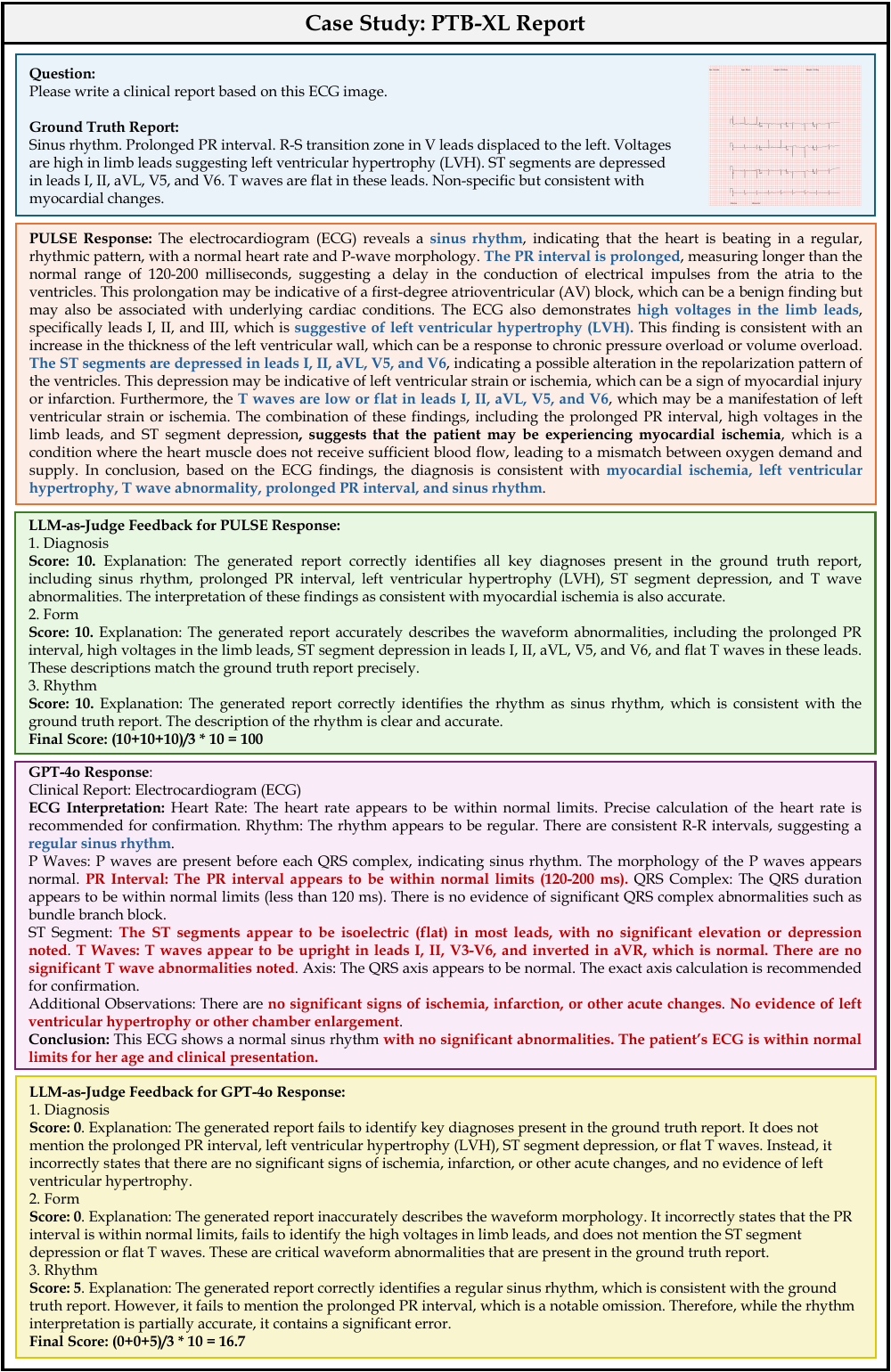}
    \caption{Comparison of model outputs on ECG report generation task (Example 1). \textcolor{blue}{Blue} indicates correct information, while \textcolor{red}{red} highlights errors. Our model's output fully aligns with the ground truth, with a report score of 10. In comparison, GPT-4's report, though structurally sound, contains notable inaccuracies despite its initial appearance of relevance.}
\label{fig:test_example_report_10}
\end{figure*}

\begin{figure*}[!t]
    \centering
\includegraphics[width=0.95\textwidth]{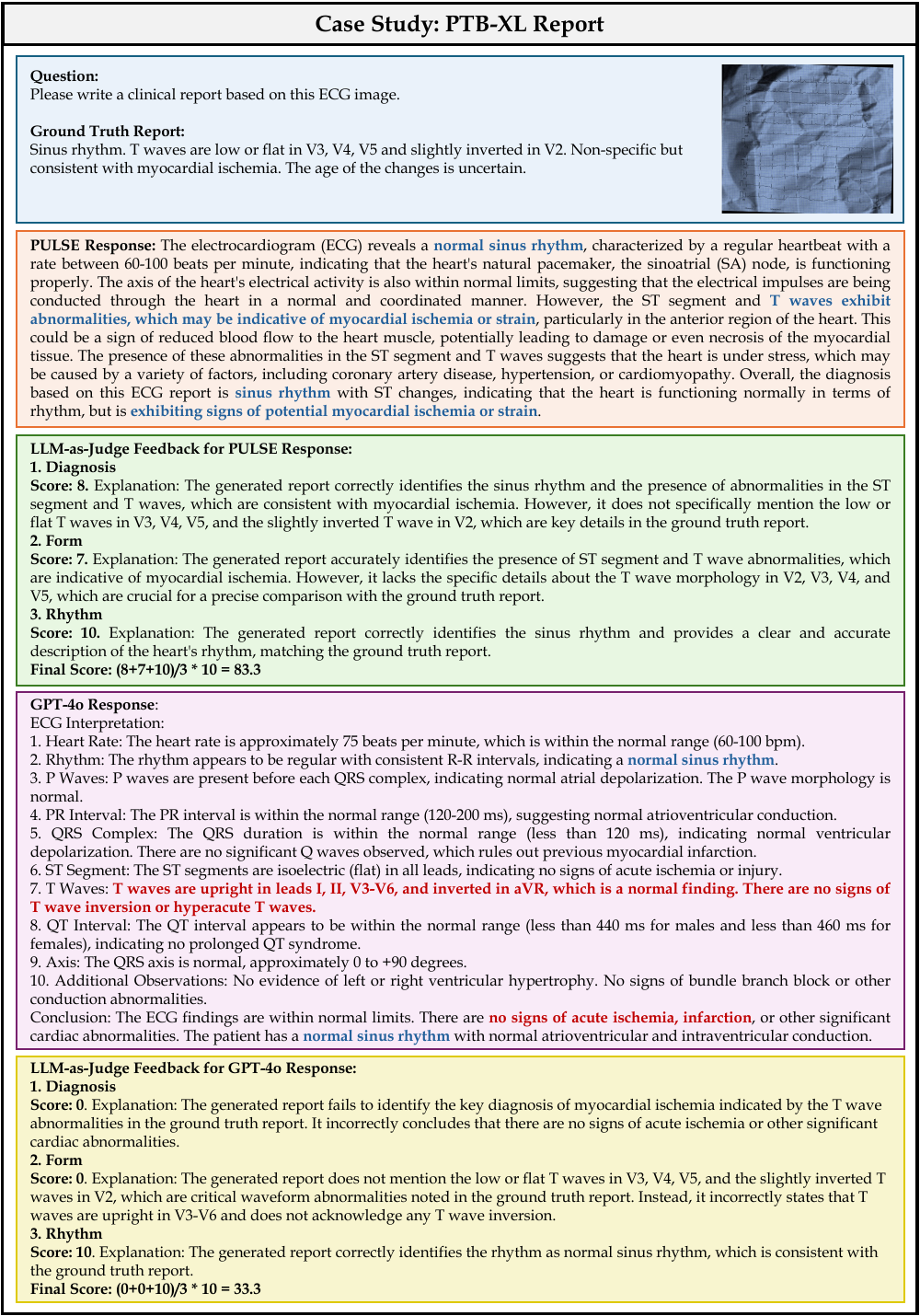}
    \caption{Comparison of model outputs on ECG report generation task (Example 2). \textcolor{blue}{Blue} indicates correct information, while \textcolor{red}{red} highlights errors. Our model's output mostly aligns with the ground truth, achieving a report score of 83.3. In comparison, GPT-4's output correctly identifies only the ECG rhythm, omitting most other key details.}
\label{fig:test_example_report_2693}
\end{figure*}

\begin{figure*}[!t]
    \centering
\includegraphics[width=0.95\textwidth]{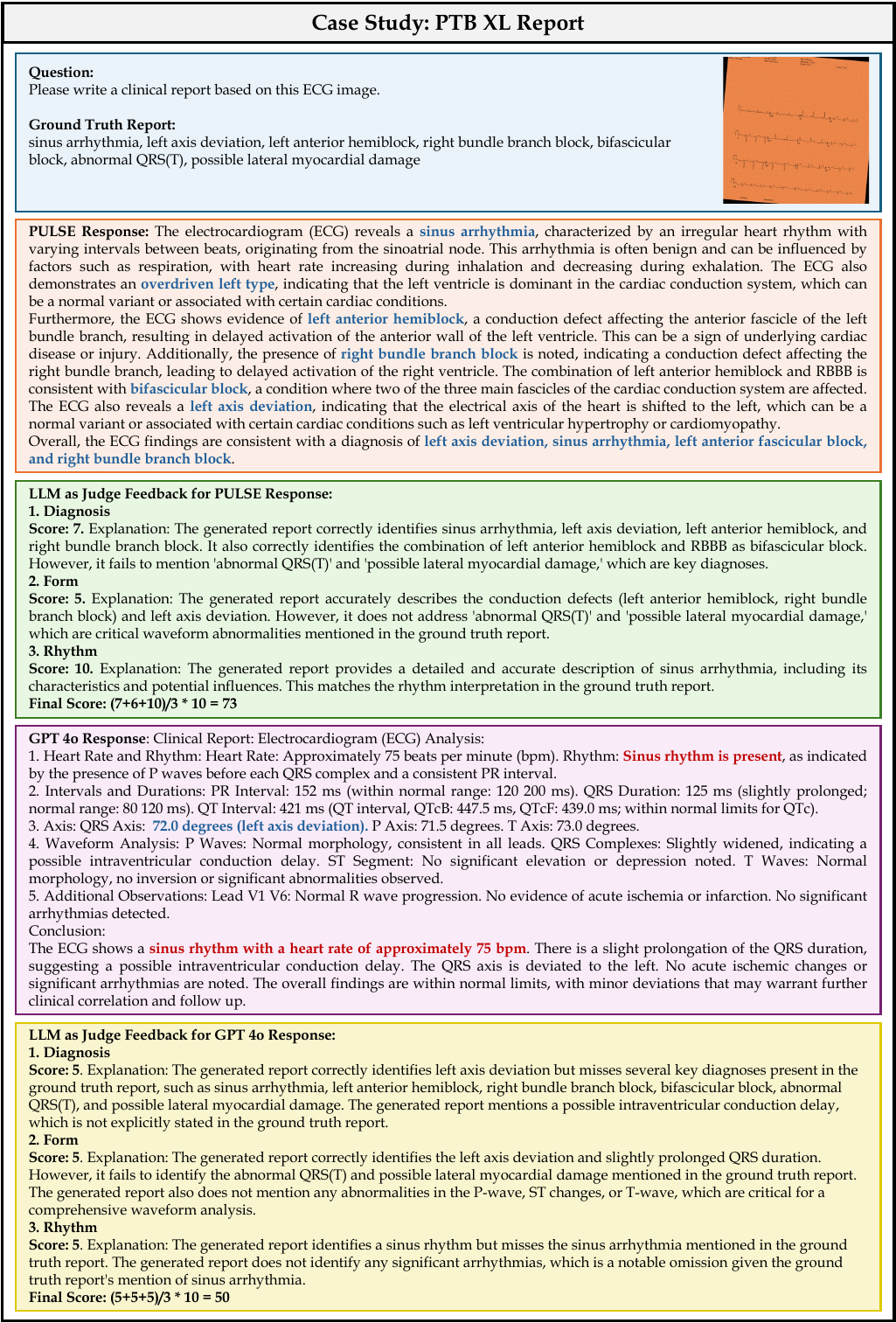}
    \caption{Comparison of model outputs on ECG report generation task (Example 3). \textcolor{blue}{Blue} indicates correct information, while \textcolor{red}{red} highlights errors. Our model's output mostly aligns with the ground truth report, achieving a report score of 73. In comparison, GPT-4's output partially aligns with the ground truth report.}
\label{fig:test_example_report_5338}
\end{figure*}

\begin{figure*}[!t]
    \centering
\includegraphics[width=0.9\textwidth]{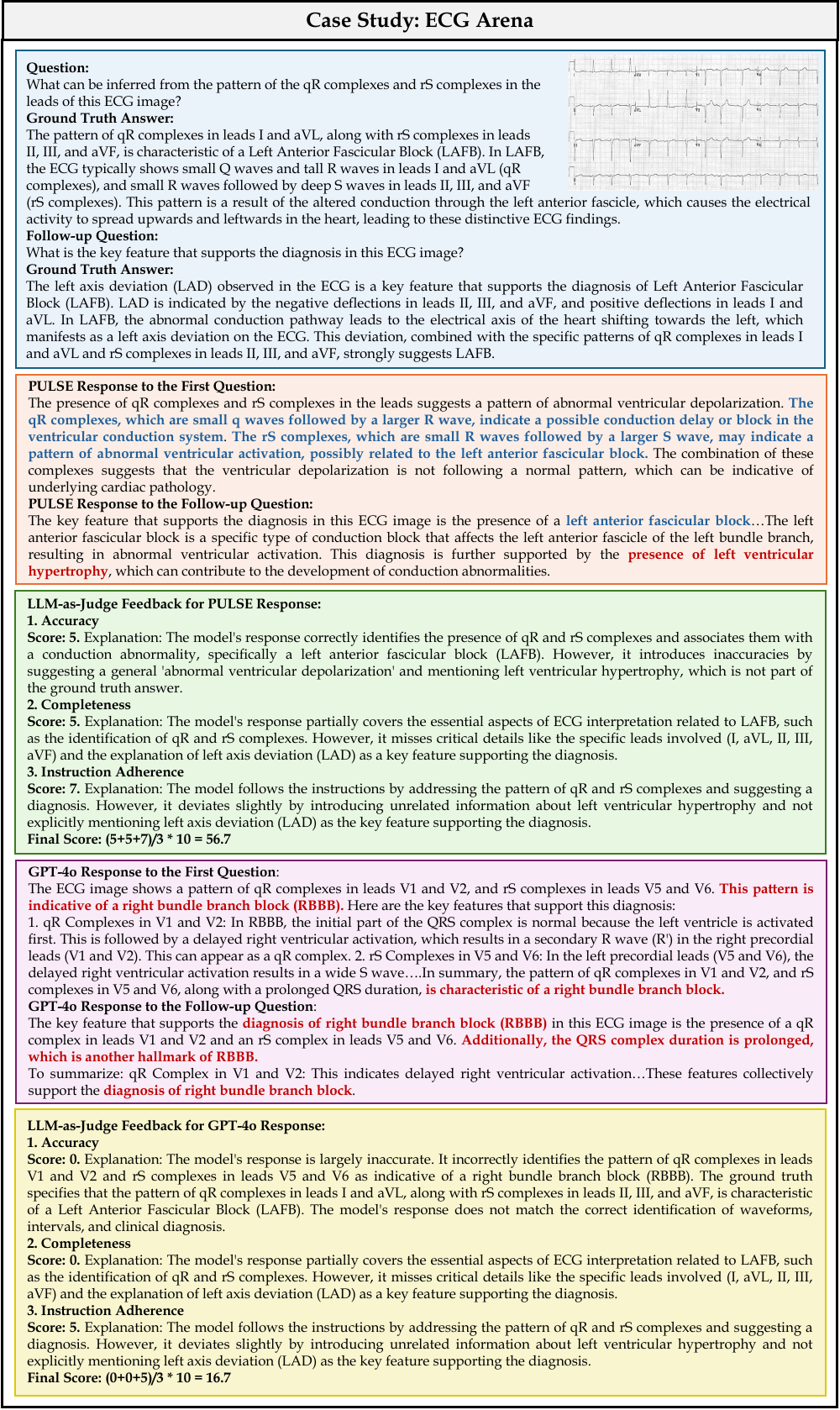}
    \caption{Comparison of model outputs on ECG Arena (Example 1). \textcolor{blue}{Blue} indicates correct information, while \textcolor{red}{red} highlights errors. Given the challenging nature of this task, our model's output partially aligns with the ground truth, and GPT-4o's output largely deviates from the reference.}
\label{fig:test_example_arena_1}
\end{figure*}

\end{document}